\documentclass[reprint,prd,nofootinbib,aps,superscriptaddress,preprintnumbers]{revtex4-2}
\usepackage[x11names]{xcolor}
\usepackage[colorlinks=true,linkcolor=blue,citecolor=Blue4,urlcolor=Blue4]{hyperref}
\usepackage{amsmath,amssymb,amsfonts,subfig,graphicx,nicefrac}
\usepackage[T1]{fontenc}
\usepackage{orcidlink}
\usepackage{cancel}
\usepackage{mathrsfs}
\usepackage[normalem]{ulem}
\usepackage{ifthen,nccmath}
\usepackage{xparse}
\usepackage{eqparbox}
\newcommand{\eqlhs}[1]{\eqmakebox[lhscol][r]{$#1$}}

\def\phi{\varphi}

\def\epsilon{\varepsilon}
\def\d{\mathrm{d}}
\def\p{\partial}
\newcommand{\Lie}[1]{\vec{\mathcal{L}}_{\vec{#1}}}
\renewcommand{\vec}[1]{\boldsymbol{#1}}

\newcommand{\DeltaPlus}{R_{\mathrm{p}}R_{\mathrm{m}}}
\newcommand{\GHPrho}{\varrho}
\newcommand{\weyl}[1]{\dot{\psi}_{#1}}

\newcommand{\ka}{a}
\newcommand{\Krho}{\rho}
\newcommand{\Krhocc}{\bar{\rho}}

\newcommand{\mass}{m}
\newcommand{\KVxi}{{\xi_{(t)}}}

\newcommand{\thorn}{\text{\th}}

\allowdisplaybreaks 

\usepackage{xstring}
\NewDocumentCommand{\cst}{m}
  {%
    \ensuremath
      {%
        \IfStrEqCase{#1}
          {%
            {-2,1}{\mathfrak{A}_1}%
            {-2,2}{\mathfrak{A}_2}%
            {-1,1}{\mathfrak{B}_1}%
            {-1,2}{\mathfrak{B}_2}%
            {0,1}{\mathfrak{C}_1}%
            {0,2}{\mathfrak{C}_2}%
            {1,1}{\mathfrak{D}_1}%
            {1,2}{\mathfrak{D}_2}%
          }[chyyyba_{#1}]
      }%
  }

\begin{document}
\preprint{YITP-26-106}
\title{Reconstruction of the Weyl\texorpdfstring{\,--\,}{ -- }Lewis\texorpdfstring{\,--\,}{ -- }Papapetrou metric for stationary and axially symmetric gravitational perturbations of a Kerr black hole}

\author{David Kofro\texorpdfstring{\v{n}}{n}\, \protect\orcidlink{0000-0002-0278-7009}}
\email{d.kofron@gmail.com}
\affiliation{Institute of Theoretical Physics, Faculty of Mathematics and Physics, Charles University, V Hole\v{s}ovi\v{c}k\'{a}ch 2, 180\,00 Prague 8, Czech Republic}
\affiliation{Astronomical Institute of the Czech Academy of Sciences, Boční II 1401/1a, 141 00 Prague, Czech Republic}

\author{Petr Kotla\v{r}\'ik\, \protect\orcidlink{0000-0002-9228-0788}}
\email{kotlarik.petr@gmail.com}
\affiliation{Astronomical Institute of the Czech Academy of Sciences, Boční II 1401/1a, 141 00 Prague, Czech Republic}
\affiliation{Center for Gravitational Physics and Quantum Information, Yukawa Institute for Theoretical Physics, Kyoto University, 606-8502, Kyoto, Japan}

\keywords{Black hole perturbations, Kerr black hole, Debye potentials, Hertz potential, Metric reconstruction}

\begin{abstract}
    The study of black hole perturbations typically follows two main approaches: the direct perturbation of the metric, or the perturbation within the Newman\,--\,Penrose (NP) or Geroch\,--\,Held\,--\,Penrose (GHP) formalism. In the latter case, a reconstruction procedure, such as the Chrzanowski\,--\,Cohen\,--\,Kegeles (CCK) method based on the Debye (Hertz) potential, is required to obtain the corresponding metric perturbation. However, the reconstructed metric is then expressed in the radiation gauge, which is not always optimal. In this paper, we analyze stationary and axially symmetric perturbations of the Kerr black hole within both frameworks. Focusing on the vacuum part of the spacetime (outside the sources), we derive an explicit gauge transformation between the averaged radiation gauge and the gauge in which the metric takes its standard Weyl\,--\,Lewis\,--\,Papapetrou (WLP) form, and we express the linearized WLP metric functions directly in terms of the Debye potential. We further discuss perturbations of the Kerr black hole towards general type D spacetimes, and analyze the mass and angular momentum perturbations in more detail. Finally, we illustrate the procedure on two examples beyond the type D class: a perturbation of the Schwarzschild black hole by a thin disk, and a perturbation of the Kerr black hole by a rotating particle.
\end{abstract}
\maketitle

\section{Introduction}
\label{sec:intro}
Black holes, one of the most important consequences of Einstein's general relativity, have been intensively studied ever since 1915, when Karl Schwarzschild discovered a solution to Einstein's equations representing the simplest black hole: a unique, vacuum, spherically symmetric black hole described by a single parameter, its mass.

After more than a century of theoretical and experimental development, black hole candidates are now the subject of intensive observational effort, including gravitational waves from black hole inspirals and direct observations of black hole shadows.  

Although typically modeled by the Kerr metric, representing an isolated, stationary, vacuum black hole, real astrophysical black holes are rarely isolated. Through mutual gravitational interaction with surrounding matter, structures such as disks, rings, or tori form around them. For a system in stationary equilibrium, where accretion onto the black hole is negligible, it is natural to assume that the spacetime is \textit{circular}, meaning that it is stationary, axially symmetric, and orthogonally transitive. In the vacuum region (outside the sources), such a spacetime can be described by the Weyl\,--\,Lewis\,--\,Papapetrou (WLP) metric~\cite{stephani_exact_2003}:
\begin{multline}
    \d s^2 = e^{2\nu}\left(\d t-\Omega\d\phi\right)^2 - \GHPrho^2 e^{-2\nu}\d\phi^2 - \\ 
    - e^{2(\lambda-\nu)}\left(\d\GHPrho^2 +\d z^2\right) \,,
\label{eq:WLP}
\end{multline}
where all the metric functions depend only on the so-called Weyl coordinates $(\GHPrho,z)$~\cite{weyl1917,stephani_exact_2003}.
The function $\nu$ is an analog of the Newtonian potential, $\Omega$ characterizes rotational dragging, and $\lambda$ describes deformations of the meridional sections. The functions $\nu$ and $\Omega$ can be combined into the the complex Ernst potential, which reduces the field equations to a single partial differential equation.

Apart from numerical simulations, one may attempt to solve the problem analytically. Even though the Einstein equations for circular spacetimes are fully integrable, exact solutions have been found only in a few idealized cases (see references below). However, given that the gravitational field of a black hole is assumed to dominate in most astrophysical systems, an approximate approach, such as black hole perturbation theory, can be employed. We may distinguish the following categories of analytical solutions:

\paragraph{Static exact solutions} In cases where the net rotation in the spacetime is negligible, the effects of rotational dragging can be neglected by setting $\Omega = 0$. This simplification reduces the field equations to a linear Laplace equation for $\nu$ and a quadrature for $\lambda$. The first exact solution describing a Schwarzschild black hole with a~static thin disk (inverted Morgan\,--\,Morgan disk \cite{morgan1969}) was considered by Lemos \& Letelier~\cite{lemos1994} and further studied in Ref.~\cite{semerak2003}. In Ref.~\cite{kofron2023}, we extended the Morgan\,--\,Morgan family to more general density profiles and completed the metric by explicitly integrating the second metric function,~$\lambda$. Semerák~\cite{semerak2004} derived another solution for the potential corresponding to disks with power-law density profiles. More recently, in Ref.~\cite{kotlarik2022}, we revisited this topic by expressing the results in closed form; the second metric function, $\lambda$, was not obtained explicitly in that case. There are two other explicit analytical solutions for a black hole encircled by a disk: in Ref.~\cite{kotlarik_black_2022}, we constructed the metric by combining the well-known Kuzmin--Toomre disks, and in Ref. \cite{vieira2020}, Vieira obtained a solution by cutting and reflecting the spacetime composed of a chain of Schwarzschild black holes. Klein~\cite{klein1997} used the soliton generation technique to study, among others, an isochrone disk around a Schwarzschild black hole, although an explicit close form of the metric was not obtained.

\paragraph{Stationary exact solutions} The stationary case is considerably more complex. However, during the 1990s, powerful generation techniques for solving the Ernst equation were developed. These techniques led to the famous Neugebauer\,--\,Meinel solution~\cite{neugebauer_einsteinian_1993,neugebauer_general_1995}, which describes a rigidly rotating thin disk of dust. Although several attempts have been made to apply these techniques to combine the gravitational fields of a black hole and a rotating disk, the explicit solutions found so far suffer from unphysical features (see, e.g., Refs.~\cite{zellerin2000,semerak2002a,klein2003,lenells2011}; cf. also Refs~\cite{klein2005,meinel2008a}).

\paragraph{Stationary perturbations} In most real astrophysical scenarios, the gravitational field of a black hole is likely to dominate that of the surrounding matter. We can therefore simplify the problem by linearizing Einstein's equations around the known black hole metric. The problem of a rotating ring around a Schwarzschild black hole was initially studied by Will~\cite{will1974,will1975} and was more recently generalized to the case of a finite disk by Čížek \& Semerák~\cite{cizek_perturbation_2017}. A further generalization to the Kerr black hole, or to a disk with a nonconstant density profile, is unlikely to be feasible analytically within the direct metric perturbation approach. On the other hand, there exists a well-established framework for black hole perturbation theory: the Newman\,--\,Penrose (NP) or the Geroch\,--\,Held\,--\,Penrose (GHP) formalism. In particular, a single complex function, the so-called Debye potential, can describe test fields of arbitrary spin, including linear perturbations of the Kerr black hole. This paper aims to provide a link between both approaches.

In his groundbreaking work, Teukolsky~\cite{Teukolsky1972,Teukolsky1973} formulated perturbation theory for the Kerr black hole in the NP formalism~\cite{Newman1962,penrose_spinors_nodate}, showing that the two NP components of a given field with the highest absolute value of spin weight satisfy a decoupled second-order partial differential equation, the so-called Teukolsky Master Equation (TME). Furthermore, he proved that the TMEs are separable in Boyer\,--\,Lindquist (BL) coordinates.

To obtain all the NP field components of arbitrary spin on type D backgrounds, it is useful to employ the Debye potential formalism. The Debye potential is a single complex scalar function that corresponds to the only independent component of Hertz potential (in a particular gauge) and is sufficient to describe the entire test field. This approach was introduced into general relativity by Chrzanowski~\cite{Chrzanowski1975}, and Cohen and Kegeles~\cite{Cohen1975}, later elaborated in Refs.~\cite{Wald1978,Kegeles1979}, and more recently generalized and clarified in terms of fundamental spinor operators by Aksteiner, Andersson and B\"ackdahl in Refs.~\cite{Andersson2015,Aksteiner2019}.

The Debye potentials were used by Linet in the late 1970s~\cite{Linet1977,Linet1979} to construct the electromagnetic field of a stationary and axisymmetric source on a Kerr background. Linet showed that in circular spacetimes the TME and the equation for the Debye potential reduce to a (flat) Laplace equation in $2s+3$ dimensions. From the generalized axially symmetric potential (GASP) theory it then follows that one need only find a solution on the symmetry axis and obtain the solution \textit{everywhere} by a single quadrature. We recently pursued this line of research further and provided the Green function for electromagnetic perturbations on the Kerr background in closed form~\cite{kofron_debye_2022}; the procedure can be straightforwardly generalized to gravitational perturbations.

However, the physical interpretation of gravitational perturbations through the Weyl scalars is challenging. A~more effective approach is to obtain the metric perturbation in a symmetry-adapted gauge, such as the WLP metric. In this paper, we present a general framework for reconstructing the perturbed WLP metric from the Debye potential.

The paper is organized as follows. In Sec.~\ref{sec:WLP}, we briefly review the WLP metric and the corresponding form of the vacuum Einstein equations. Section~\ref{sec:Kerr} provides a concise summary of the Kerr solution. In Sec.~\ref{sec:Debye}, we introduce the Debye potential for gravitational perturbations using both the standard NP and GHP frameworks~\cite{Newman1962,Geroch1973,penrose_spinors_nodate,Andersson2015}. Accordingly, we use spin coefficients $\kappa,\sigma,\GHPrho,\tau$ and the GHP derivatives $\thorn,\eth$ (and their primed counterparts). The symbol $\GHPrho$ also denotes the radial Weyl coordinate, but no confusion should arise. Section~\ref{sec:Calibration} presents the main results of this paper: an explicit gauge transformation, and expressions for the linear perturbation of the WLP metric functions in terms of the Debye potential. 
Next, pure type D vacuum perturbations within the Debye formalism are discussed in Sec.~\ref{sec:typeDpert}, with a particular focus on the mass and angular momentum perturbations and the cosmic string. That section follows up on and generalizes discussions of Wald~\cite{wald1973:perturbationsKerr}, Keidl~\cite{Keidl2007} and van de Meent~\cite{vandemeent2017:massangular}. Finally, in Sec.~\ref{sec:Examples}, we work out two explicit examples of the approach: a disk around a  static  black hole, and a rotating point particle located on the axis of the Kerr black hole, before concluding in Sec.~\ref{sec:Conclusions}.

The paper also contains four appendices. Appendix~\ref{app:Kerr} lists the NP quantities of the Kerr black hole with respect to the null tetrad used, and Appendix~\ref{sec:xyz} gives the coordinate expression of some extensively used functions. Appendix~\ref{app:csce} enumerates all the gauge equations, and in App.~\ref{app:Killing} we derive some NP/GHP identities for non-accelerating spacetimes of algebraic type~D, which provide an independent derivation of presented results (one could, of course, have worked directly in coordinates from the very beginning).

We also provide a \emph{Wolfram Mathematica}\textsuperscript{\copyright} notebook~\cite{10.5281/zenodo.21699521}, which contains the main results of this paper and several worked examples.

We use geometrized units in which the speed of light $c$ and the gravitational constant $G$ equal unity. The metric signature is $(+,-,-,-)$, and the partial derivatives are denoted either by prefix notation $\frac{\p f}{\p x}$ or postfix notation~$f_{,x}$.

\section{Weyl\texorpdfstring{\,--\,}{ -- }Lewis\texorpdfstring{\,--\,}{ -- }Papapetrou class of metrics}
\label{sec:WLP}

The stationary and axially symmetric vacuum Einstein equations for the WLP metric~(\ref{eq:WLP}) consist of two Poisson-type coupled equations
\begin{align}
    \Delta_0 \nu +\frac{e^{4\nu}}{2\GHPrho^2}\,\left(\nabla \Omega\right)^2 &=0\,, \label{eq:psi} \\
    \Delta_{-1} \Omega + 4\nabla\nu\cdot\nabla\Omega &=0\,, \label{eq:Omega}
\end{align}
where $\nabla$ and $\nabla\cdot$ denote the standard gradient and divergence operators (in an auxiliary flat space), and $\Delta_s$ is the generalized $(2s + 3)$-dimensional Laplace operator in cylindrical coordinates, acting on axially symmetric function,
\begin{equation}
    \Delta_s = \frac{\p^2}{\p \GHPrho^2}+\frac{2s+1}{\GHPrho}\frac{\p}{\p\GHPrho}+\frac{\p^2}{\p z^2} \,.
\end{equation}
The Einstein equations are completed by a pair of equations for the metric function $\lambda$:
\begin{align}
    \lambda_{,\GHPrho} &= \GHPrho\left(\nu_{,\GHPrho}^2-\nu_{,z}^2\right)-\frac{e^{4\nu}}{4\GHPrho}\left(\Omega_{,\GHPrho}^2-\Omega_{,z}^2\right) ,\label{eq:lambdar}\\
    \lambda_{,z} &= 2\GHPrho\nu_{,\GHPrho}\nu_{,z}-\frac{e^{4\nu}}{2\GHPrho}\,\Omega_{,\GHPrho}\Omega_{,z}  \,.\label{eq:lambdaz}  
\end{align}

The first pair of equations, (\ref{eq:psi})\,--\,(\ref{eq:Omega}), can be combined into a single Ernst equation for the complex Ernst potential~\cite{ernst_new_1968,ernst_complex_1974}. Although the Ernst equation has been shown to be fully integrable, and several classes of exact solutions have been derived by generation techniques, many of them lack a clear physical interpretation.

In the case of vanishing rotation ($\Omega=0$), the Ernst equation reduces to the Laplace equation. Consequently, any axially symmetric solution known from Newtonian theory is also a solution in general relativity, provided that the corresponding second metric function $\lambda$ is determined by integrating Eqs.~(\ref{eq:lambdar}) and~(\ref{eq:lambdaz}) with $\Omega = 0$. For non-vanishing rotation ($\Omega \neq 0$), the full system of Ernst equations and the equations for $\lambda$ must be solved. For a brief review of the known solutions, see Sec.~\ref{sec:intro}.

\section{Kerr black hole}
\label{sec:Kerr}
The rotating black hole --- one of the most astrophysically significant solutions of the vacuum Einstein field equations --- was discovered in 1963 by Roy Kerr~\cite{Kerr1963} (cf. the recent historical reviews in Refs.~\cite{Wiltshire2009,Teukolsky2015}).

The metric in Boyer\,--\,Lindquist (BL) coordinates $(t, r, \theta, \varphi)$ is given by
\begin{multline}
\vec{\d} s^2 = \frac{\Delta}{\Sigma} \left( \vec{\d} t -\ka\sin^2\theta\,\vec{\d}\phi \right)^2 -\frac{\Sigma}{\Delta}\,\vec{\d} r^2 \\
- \Sigma \, \vec{\d}\theta^2  
-\frac{\sin^2\theta}{\Sigma}\Bigl[ \left( \ka^2+r^2 \right)\vec{\d}\phi - \ka\,\vec{\d} t \Bigr]^2\,,
\label{eq:KerrMetric}
\end{multline}
with the following standard definitions
\begin{align}
\Delta&=r^2-2\mass r+\ka^2=(r-r_\mathrm{p})(r-r_\mathrm{m})\,,\\
\rho&=r-i\ka\cos\theta\,,\label{eq:Krho}\\
\Sigma&=\rho\bar{\rho}=r^2+\ka^2\cos^2\theta\,.
\end{align} 
For later reference, we also define
\begin{align}
    \Upsilon = -\frac{\Sigma}{\sin^2\theta}\, g_{\phi\phi}=\Sigma\Delta+2\mass(r^2+\ka^2)\,. \label{eq:UpsilonDef}
\end{align}
The parameters have the following interpretation: $\mass$ is the mass of the black hole and $\mass\ka$ its angular momentum, while $r_\mathrm{p}$ and $r_\mathrm{m}$ are the position of the outer and the inner black hole horizon, respectively,
\begin{equation}
    r_{\mathrm{p},\mathrm{m}} =\mass \pm \sqrt{\mass^2 - a^2} \,.
\end{equation}
We will also frequently use the parameter $\beta$, defined as\footnote{Of the parameters $r_\mathrm{p}$, $r_\mathrm{m}$, $\mass$,$\ka$, and $\beta$, only two are independent.}
\begin{equation}
\beta=\sqrt{\mass^2-\ka^2}=(r_\mathrm{p}-r_\mathrm{m})/2\,.
\end{equation}

When working in the NP formalism, we adopt a slightly modified version of the Kinnersley tetrad~\cite{kinnersley1969}, denoted by $(\vec{l},\,\vec{n},\,\vec{m},\,\bar{\vec{m}})$. Compared with the standard textbook form, our tetrad is boosted in the $\vec{l}\vec{n}$ plane by a factor of $\sqrt{2}$, which makes the resulting expressions in terms of the Debye potential ``more symmetric''. Explicitly:
\begin{equation}
\begin{alignedat}{1}
\vec{l} &= \frac{1}{\sqrt{2}\,\Delta}\left[ \left( r^2+\ka^2 \right)\vec{\p_t}+\Delta\,\vec{\p_r} + \ka\,\vec{\p_\phi} \right]\,, \\
\vec{n} &= \frac{1}{\sqrt{2}\,\Sigma}\left[ \left( r^2+\ka^2 \right)\vec{\p_t}-\Delta\,\vec{\p_r} + \ka\,\vec{\p_\phi} \right]\,, \\
\vec{m} &= \frac{1}{\sqrt{2}\,\bar{\rho}}\bigl( i\ka\sin\theta\,\vec{\p_t}+\vec{\p_\theta} +i\csc\theta\,\vec{\p_\phi} \bigr)\,,\\
\vec{\bar{m}} &= \frac{1}{\sqrt{2}\,\rho}\bigl( -i\ka\sin\theta\,\vec{\p_t}+\vec{\p_\theta} -i\csc\theta\,\vec{\p_\phi} \bigr)\,.
\end{alignedat}
\label{eq:NPtetrad}
\end{equation}

In order to transform the Kerr metric from BL to Weyl coordinates, we define the functions $R_\mathrm{p}$, $R_\mathrm{m}$ as
\begin{align}
    R_\mathrm{p} &= \sqrt{\GHPrho^2+(z-\beta)^2}\,, &
    R_\mathrm{m} &= \sqrt{\GHPrho^2+(z+\beta)^2}\,. \label{eq:RpRm}
\end{align}
The transformation to Weyl coordinates then reads
\begin{align}
    r &= \mass+\frac{R_\mathrm{p}+R_\mathrm{m}}{2}\,,&
    \theta &=\arccos\left(-\frac{R_\mathrm{p}-R_\mathrm{m}}{2\beta}\right)\,,
\label{eq:toW}
\end{align}
and the inverse of the transformation~(\ref{eq:toW}) is
\begin{align}
    z&=\nicefrac{1}{2}\,\Delta'(r)\cos\theta\,, &
    \GHPrho&=\sqrt{\Delta}\,\sin\theta \,.
\end{align}

The Kerr metric (\ref{eq:KerrMetric}) in the WLP form (\ref{eq:WLP}) has the following metric functions:\footnote{We add the subscript ``$_0$'' to indicate that these are the background values for the subsequent linearization.}
\begin{align}
\nu_0 &= \frac{1}{2}\,\ln\left[\frac%
{\left(R_\mathrm{p}+R_\mathrm{m}\right)^2-4\mass^2+\frac{a^2}{\beta^2}\left(R_\mathrm{p}-R_\mathrm{m}\right)^2}%
{\left(R_\mathrm{p}+R_\mathrm{m}+2\mass\right)^2+\frac{a^2}{\beta^2}\left(R_\mathrm{p}-R_\mathrm{m}\right)^2}\right] , \label{eq:Knu}\\
\lambda_0 &= \frac{1}{2}\,\ln\left[\frac%
{\left(R_\mathrm{p}+R_\mathrm{m}\right)^2-4\mass^2+\frac{a^2}{\beta^2}\left(R_\mathrm{p}-R_\mathrm{m}\right)^2}%
{4R_\mathrm{p}R_\mathrm{m}}\right] , \label{eq:Klambda}\\
\Omega_0 &= \frac{a\mass}{\beta^2}\,\frac%
{\left(R_\mathrm{p}+R_\mathrm{m}+2\mass\right)\left[\left(R_\mathrm{p}-R_\mathrm{m}\right)^2-4\beta^2\right]}%
{\left(R_\mathrm{p}+R_\mathrm{m}\right)^2-4\mass^2+\frac{a^2}{\beta^2}\left(R_\mathrm{p}-R_\mathrm{m}\right)^2}\,. \label{eq:KOmega}
\end{align}

We close this section by introducing some GHP operators associated with the symmetries of the Kerr background. The Kerr spacetime admits the stationary Killing vector $\vec{\KVxi}=\vec{\p_t}$, the axial Killing vector $\vec{\eta}=\vec{\p_\phi}$ and the vector $\vec{\zeta} =a^2\vec{\p_t}+a\vec{\p_\phi}$), which arises from the Killing tensor of the Kerr background. These symmetries give rise to GHP operators that, acting on a GHP scalar $\psi_{[p,q]}$ of arbitrary weights $[p,q]$, commute with all the GHP derivatives:
\begin{align}
    \vec{K_\KVxi} &= -\kappa_1 \left(
    -\GHPrho'\thorn+\GHPrho\thorn'+\tau'\eth-\tau\eth'\right) \\
    & \qquad -\frac{p}{2}\,\kappa_1\psi_2-\frac{q}{2}\bar{\kappa}_{1'}\,\bar{\psi}_2
    \,, \nonumber\\
    \vec{K_\zeta} &= -\frac{1}{4}\kappa_1 \Bigl[
    \left(\kappa_1-\bar{\kappa}_{1'}\right)^2\left(\GHPrho'\thorn-\GHPrho\thorn'\right) \\
    &\qquad-\left(\kappa_1+\bar{\kappa}_{1'}\right)^2\left(\tau'\eth-\tau\eth'\right)\Bigr]+\frac{p}{8}\kappa_1\mathcal{K}+\frac{q}{8}\bar{\kappa}_{1'}\bar{\mathcal{K}}\,, \nonumber\\
    \mathcal{K}&=\psi_2\left(\kappa_1^2+\bar{\kappa}_{1'}^2\right) - 2\bar{\psi}_2\, \bar{\kappa}_{1'}^2\\
    &\qquad+4\kappa_1\bar{\kappa}_{1'}\left[ 
    \GHPrho\left( \GHPrho'-\bar{\GHPrho}'\right)+\tau'\left(\tau-\bar{\tau}'\right) \nonumber
    \right]\,.
\end{align}
A detailed discussion of Killing spinors, vectors and tensors, together with proofs of some necessary identities, can be found in Appendix~\ref{app:Killing}. These results are not essential for our calculations, but they can be used to prove the relation between the two Debye potentials, Eq.~(\ref{eq:DebMaxFromDebMin}) of Sec.~\ref{sec:BlackHolePerturbations}, not only in coordinates but also in a coordinate-independent fashion within the GHP formalism.

\section{Black hole perturbations}
\label{sec:BlackHolePerturbations}
Let us consider a linear perturbation of a black hole controlled by a small parameter $\epsilon \ll 1$,
\begin{equation}
\vec{g}(\epsilon)_{ab} = \vec{g}_{ab}+\epsilon \vec{h}_{ab} \,,
\end{equation}
where $\vec{g}_{ab}$ is the background metric of the Kerr black hole, and $\vec{h}_{ab}$ encodes the linear metric perturbation. This induces perturbations of the NP tetrad vectors and of all NP (GHP) quantities. Background (unperturbed) quantities are not be denoted in any special way, while their perturbations are denoted by a dot, e.g., $\psi_0 \rightarrow \psi_0 + \epsilon \dot{\psi}_0$.

\subsection{Debye potentials and the metric reconstruction}
\label{sec:Debye}
Gravitational perturbations of vacuum type D spacetimes can be determined from a single complex scalar field --- a particular component of the Hertz potential (see Refs~\cite{Deadman2011,Aksteiner2019,kofron2020}; cf. a comprehensive review of perturbation theory in Ref.~\cite{pound2022}). Two different Debye potentials can be introduced, depending on the boundary conditions imposed on the horizon: they correspond to (a) the outgoing radiation gauge (ORG), and (b) the ingoing radiation gauge (IRG). In what follows, we consider only stationary and axially symmetric vacuum perturbations of the Kerr black hole and detail the CCK metric reconstruction procedure in this case.

\subsubsection{Outgoing radiation gauge}
By defining the spinorial Hertz potential as $\chi_{ABCD}=\chi_{[4,0]}\iota_A\iota_B\iota_C\iota_D$, we introduce the Debye potential $\chi_{[4,0]}$, which is a GHP scalar of weights $[4,0]$. It satisfies the following Debye equation:
\begin{equation}
\bigl[ \left( \thorn -\bar{\GHPrho} \right)\left( \thorn' +3\GHPrho' \right)  
- \left( \eth-\bar{\tau}'\right)\left( \eth'+3\tau' \right) 
-3\psi_2 \bigr] \chi_{[4,0]}  = 0 \label{eq:GHPDebMax}  \,.
\end{equation}
The resulting metric perturbation, reconstructed from this Debye potential, is in the ORG and reads
\begin{multline}
\vec{h}^{(\mathrm{out})}_{ab}=
    \left(X_\mathrm{o}+\bar{X}_\mathrm{o}\right)\vec{n}_a\vec{n}_b
    +Y_\mathrm{o}\vec{m}_a\vec{m}_b+\bar{Y}_\mathrm{o}\vec{\bar{m}}_a\vec{\bar{m}}_b\\
    -2Z_\mathrm{o}\vec{n}_{(a}\vec{m}_{b)}-2\bar{Z}_\mathrm{o}\vec{n}_{(a}\vec{\bar{m}}_{b)}\,, \label{eq:hORG}
\end{multline}
where
\begin{align}
X_\mathrm{o} &= \left(\eth'\eth'+2\tau'\eth'\right)\chi_{[4,0]} \,, \\
Y_\mathrm{o} &= \left(\thorn'\thorn'+2\bar{\GHPrho}'\thorn'\right)\bar{\chi}_{[0,4]} \,,\\
Z_\mathrm{o} &= \left(\thorn'\eth+\left(\tau+\bar{\tau}'\right)\thorn'+\bar{\GHPrho}'\eth\right)\bar{\chi}_{[0,4]} \,.
\end{align}
This metric perturbation clearly satisfies $\vec{h}^{(\mathrm{out})}_{ab} \vec{n}^a = 0$ and $\vec{g}^{ab} \vec{h}^{(\mathrm{out})}_{ab} = 0$.

For the perturbations $\dot{\psi}_0$ and $\dot{\psi}_4$ of the Weyl scalars, we have 
\begin{align}
    2\dot{\psi}_0 &= - \eth\eth\eth\eth\bar{\chi}_{[0,4]}-3\psi_2\kappa_1^{-1}\vec{K_\KVxi}\chi_{[4,0]}\,, \label{eq:psi0fromDebMax}\\
    2\dot{\psi}_4 &= -\thorn'\thorn'\thorn'\thorn' \bar{\chi}_{[0,4]} \label{eq:psi4fromDebMax} \,.
\end{align}
All the other components of the perturbed Weyl tensor and the perturbations of the spin coefficients can be found in Refs.~\cite{Deadman2011,Aksteiner2019} (note that some of the older literature may contain typographical errors).

\subsubsection{Ingoing radiation gauge}
Defining the spinorial Hertz potential in the complementary way as $\chi_{ABCD}=\chi_{[-4,0]}o_Ao_Bo_Co_D$, we introduce a different Debye potential $\chi_{[-4,0]}$, which is a GHP scalar of weights $[-4,0]$. The corresponding Debye equation, the primed version of Eq. (\ref{eq:GHPDebMax}), reads
\begin{equation}
\bigl[ \left( \thorn' -\bar{\GHPrho}' \right)\left( \thorn +3\GHPrho \right)  
- \left( \eth'-\bar{\tau}\right)\left( \eth+3\tau \right) 
-3\psi_2 \bigr] \chi_{[-4,0]}  = 0  \,. \label{eq:GHPDebMin}
\end{equation}
The metric perturbation reconstructed from this Debye potential is in the IRG and reads
\begin{multline}
\vec{h}^{(\mathrm{in})}_{ab}=
    \left(X_\mathrm{i}+\bar{X}_\mathrm{i}\right)\vec{l}_a\vec{l}_b
    +\bar{Y}_\mathrm{i}\vec{m}_a\vec{m}_b+Y_\mathrm{i}\vec{\bar{m}}_a\vec{\bar{m}}_b\\
    -2\bar{Z}_\mathrm{i}\vec{l}_{(a}\vec{m}_{b)}-2Z_\mathrm{i}\vec{l}_{(a}\vec{\bar{m}}_{b)} \,,
\end{multline}
where
\begin{align}
X_\mathrm{i} &= \left(\eth\eth+2\tau\eth\right)\chi_{[-4,0]} \,, \\
Y_\mathrm{i} &= \left(\thorn\thorn+2\bar{\GHPrho}\thorn\right)\bar{\chi}_{[0,-4]} \,,\\
Z_\mathrm{i} &= \left(\thorn\eth'+\left(\tau'+\bar{\tau}\right)\thorn+\bar{\GHPrho}\eth'\right)\bar{\chi}_{[0,-4]} \,.
\end{align}
This metric perturbation satisfies $\vec{h}^{(\mathrm{in})}_{ab} \vec{l}^a = 0$ and $\vec{g}^{ab} \vec{h}^{(\mathrm{in})}_{ab} = 0$. The perturbations of the Weyl scalars are
\begin{eqnarray}
2\weyl{0}&=& -\thorn\thorn\thorn\thorn\, \bar{\chi}_{[0,-4]}\,,\label{eq:psi0fromDebMin}\\
2\weyl{4}&=& -\eth'\eth'\eth'\eth'\, \bar{\chi}_{[0,-4]}-3\psi_2\kappa_1^{-1}\vec{K_\KVxi}\chi_{[-4,0]} \label{eq:psi4fromDebMin}\,. 
\end{eqnarray}
For stationary and axially symmetric fields, Eq.~(\ref{eq:psi0fromDebMin}) reduces in BL coordinates to
\begin{align}
\dot{\psi}_0 &= -\frac{1}{8}\frac{\p^4\,\bar{\chi}_{[0,-4]}}{\p r^4}\,.\label{eq:psi0orgBL}
\end{align}
Moreover, in both radiation gauges, the following identity holds:
\begin{align}
\dot{\psi}_4 &= \frac{\Delta^2}{\rho^4}\,\dot{\psi}_0\,.
\label{eq:psi4idBL}
\end{align}

\subsubsection{Averaging IRG and ORG}
Since the perturbations are linear, we may construct a new gauge by averaging the IRG and the ORG,
\begin{equation} 
\vec{h}_{ab} = \frac{1}{2}\left(\vec{h}^{(\mathrm{in})}_{ab}+\vec{h}^{(\mathrm{out})}_{ab}\right) \,.
\label{eq:sum}
\end{equation}
In this section, we show that this choice of gauge substantially simplifies the reconstructed metric: it makes the perturbation explicitly circular, i.e., the metric is block diagonal in $(t,\phi)$ and $(r,\theta)$. For stationary and axially symmetric perturbations we found a simple ``symmetry'' between the two Debye potentials 
\begin{equation}
\chi_{[4,0]} = \left(\frac{\bar{\GHPrho}\,\bar{\tau}'}{\bar{\GHPrho}'\,\bar{\tau}}\right)^2\chi_{[-4,0]} 
=\frac{\rho^4}{\Delta^2}\chi_{[-4,0]} \,.
\label{eq:DebMaxFromDebMin}
\end{equation}
If $\chi_{[-4,0]}$ is a solution of Eq.~(\ref{eq:GHPDebMin}) and
\begin{align}
\vec{K_\KVxi} \,\chi_{[-4,0]} &=0\,, &
\vec{K_\zeta} \, \chi_{[-4,0]} &=0\, 
\end{align}
then $\chi_{[4,0]}$ as given by Eq.~(\ref{eq:DebMaxFromDebMin}) is a solution of Eq.~(\ref{eq:GHPDebMax}) and leads to the same perturbation (in the sense of physical content, i.e. up to gauge). Using these relations, we find
\begin{align}
X_\mathrm{o} &= \frac{\Sigma^2}{\Delta^2}\, X_\mathrm{i} \,, &
Y_\mathrm{o} &= \frac{\bar{\rho}^2}{\rho^2}\, Y_\mathrm{i} \,, &
Z_\mathrm{o} &= -\frac{\bar{\rho}^2}{\Delta}\, Z_\mathrm{i}\,.
\end{align}

In BL coordinates, after some tedious algebra, the resulting metric perturbation (\ref{eq:sum}) simplifies to
\begin{equation}
\vec{h}_{ab} = \begin{pmatrix}
h_{tt} & 0 & 0 & h_{t\phi} \\
0 & h_{rr} & h_{r\theta} & 0 \\
0 & h_{r\theta} & h_{\theta\theta} & 0 \\
h_{t\phi} & 0 & 0 & h_{\phi\phi}
\end{pmatrix}\,,
\end{equation}
which is block diagonal in $(t,\phi)$ and $(r,\theta)$. In particular,
\begin{align}
2\vec{h} &=
    (x+\bar{x})\left[\left(\vec{\d}t-\ka\sin^2\theta\,\vec{\d}\phi\right)^2+\frac{\Sigma^2}{\Delta^2}\,\vec{\d} r^2\right] \nonumber\\
    &
    -(y+\bar{y})\left[\left(\ka\,\vec{\d}t-(r^2+\ka^2)\vec{\d}\phi\right)^2\sin^2\theta-\Sigma^2\vec{\d}\theta^2\right] \nonumber \\
    &-2i(z-\bar{z})\sin\theta\left[-\ka\,\vec{\d}t+(r^2+\ka^2)\vec{\d}\phi\right]\times\nonumber\\ 
    &\quad \left(\vec{\d}t-a\sin^2\theta\,\vec{\d}\phi\right)     
 -2(z+\bar{z})\frac{\Sigma^2}{\Delta}\,\vec{\d}r\,\vec{\d}\theta \,,
 \label{eq:pDeb}
\end{align}
where we have introduced\footnote{Also the symbol $z$ is overloaded but, again, no confusion is expected to arise.}
\begin{align}
x &= X_\mathrm{i} = X_\mathrm{o} \Delta^2 \Sigma^{-2} \,, \\
y &= Y_\mathrm{i}\, \rho^{-2} = Y_\mathrm{o}\, \bar{\rho}^{-2} \,, \\
z &=- Z_\mathrm{i}\, \rho^{-1} = -Z_\mathrm{o}\, \Delta \Sigma^{-1}\bar{\rho}^{-1} \,.
\end{align}
The explicit coordinate expressions of the scalars $x$, $y$, and $z$ are listed in Appendix~\ref{sec:xyz}.

\subsection{Direct circular metric perturbations}

Consider an additional circular source of gravity around the Kerr black hole, i.e., a source respecting the symmetries, namely stationarity and axial symmetry. Outside the source (in vacuum), the metric of such a system can always be expressed in the WLP form (\ref{eq:WLP}). If the spacetime is only slightly deformed, we can expand the metric functions about their Kerr values (\ref{eq:Knu})\,--\,(\ref{eq:KOmega}),
\begin{align}
\nu &\rightarrow \nu_0 + \epsilon \nu_1 \,, &
\lambda &\rightarrow \lambda_0 + \epsilon \lambda_1 \,, &
\Omega &\rightarrow \Omega_0 + \epsilon \Omega_1 \,.
\end{align}
The linear contribution to the metric then reads
\begin{align}
\vec{\tilde{h}}&=2e^{2\nu_0}\left(\vec{\d}t-\Omega_0\vec{\d}\phi\right)\left[\nu_1\vec{\d}t-\left(\Omega_0\nu_1+\Omega_1\right)\vec{\d}\phi\right] \\
&\quad -2e^{2(\lambda_0-\nu_0)}\left(\lambda_1-\nu_1\right)\left(\vec{\d}\GHPrho^2+\vec{\d}z^2\right)-2\GHPrho^2e^{-2\nu_0} \nu_1\vec{\d}\phi^2\,.\nonumber
\end{align}
The corresponding linearizations of Einstein equations (\ref{eq:psi}), (\ref{eq:Omega}), and (\ref{eq:lambdar}), (\ref{eq:lambdaz}) in Weyl coordinates read 
\begin{align}
    \Delta_0 \nu_1 &=-e^{4\nu_0}\GHPrho^{-2}\left(2\nu_1\nabla\Omega_0+\nabla\Omega_1\right)\cdot\nabla\Omega_0 \,,\label{eq:linNu}\\
    \Delta_{-1}\Omega_1&=-4\left(\nabla\nu_0\cdot\nabla\Omega_1+\nabla\nu_1\cdot\nabla\Omega_0\right) \,,\label{eq:linOmega}\\
    \lambda_{1,z}&=2\GHPrho\left(\nu_{0,\GHPrho}\nu_{1,z}+\nu_{0,z}\nu_{1,\GHPrho}\right) - \nicefrac{1}{2}\, e^{4\nu_0}\GHPrho^{-1}\times  \\
    &\quad\left(4\nu_1\Omega_{0,\GHPrho}\Omega_{0,z}+\Omega_{1,z}\Omega_{0,\GHPrho}+\Omega_{0,z}\Omega_{1,\GHPrho}\right),\nonumber\\
    \lambda_{1,\GHPrho}&=2\GHPrho\left(\nu_{0,\GHPrho}\nu_{1,\GHPrho}-\nu_{0,z}\nu_{1,z}\right)  +\nicefrac{1}{2}\, e^{4\nu_0}\GHPrho^{-1}\times  \label{eq:linGammar}\\
    &\quad\left(2\nu_1\left(\Omega_{0,z}^2-\Omega_{0,\GHPrho}^2\right)+\Omega_{1,z}\Omega_{0,z}-\Omega_{0,\GHPrho}\Omega_{1,\GHPrho}\right)\,.\nonumber
\end{align}

In BL coordinates, the metric perturbation tensor reads
\begin{equation}
\vec{\tilde{h}}_{ab} = \begin{pmatrix}
\tilde{h}_{tt} & 0 & 0 & \tilde{h}_{t\phi} \\
0 & \tilde{h}_{rr} & 0 & 0 \\
0 & 0 & \tilde{h}_{\theta\theta} & 0 \\
\tilde{h}_{t\phi} & 0 & 0 & \tilde{h}_{\phi\phi}
\end{pmatrix},
\label{eq:pWeyl}
\end{equation}
where the components are
\begin{align}
{\tilde{h}}_{tt} &= 2\frac{\Sigma_m^-}{\Sigma}\, \nu_1 \,, \label{eq:pWeyltt}\\
{\tilde{h}}_{t\phi} &= 4\frac{\ka \mass r\sin^2\theta}{\Sigma}\, \nu_1 - \frac{\Sigma_\mass^-}{\Sigma}\,\Omega_1 \,, \\
{\tilde{h}}_{\phi\phi} &= 2\sin^2\theta\,\frac{4\ka^2\mass^2r^2\sin^2\theta+\Delta\Sigma^2}{\Sigma\Sigma_m^-}\,\nu_1\nonumber\\
&\qquad\qquad- \sin^2\theta\,\frac{4\ka \mass r}{\Sigma}\, \Omega_1 \,, \\
{\tilde{h}}_{rr} &= 2\frac{\Sigma}{\Delta} \left(\nu_1-\lambda_1\right) , \\
{\tilde{h}}_{\theta\theta} &= 2\Sigma \left(\nu_1-\lambda_1\right) , \label{eq:pWeylthth}
\end{align} 
with $\Sigma_m^\pm$ defined as
\begin{align}
    \Sigma_m^\pm &\equiv \Sigma\pm 2\mass r\,. \label{eq:SigmapmDef}
\end{align}

\section{Metric reconstruction in the Weyl\texorpdfstring{\,--\,}{ -- }Lewis\texorpdfstring{\,--\,}{ -- }Papapetrou gauge}
\label{sec:Calibration}
The metric perturbation $\vec{h}$~(\ref{eq:pDeb}), obtained from the Debye potential, and $\vec{\tilde{h}}$~(\ref{eq:pWeyl}), given by the linearization of the WLP metric, are connected by a gauge transformation
\begin{equation}
\vec{h} = \vec{\tilde{h}} + \Lie{\xi}\vec{g} \,.
\label{eq:Deb-vs-WLP}
\end{equation}
We choose the following ansatz for the gauge vector~$\vec{\xi}$ in BL coordinates
\begin{equation}
\vec{\xi}=\xi^r(r,\theta) \, \vec{\p_r}+ \xi^\theta(r,\theta) \, \vec{\p_\theta}\,. \label{eq:GaugeVectorAnsatz}
\end{equation}
Eq.~(\ref{eq:Deb-vs-WLP}) thus represents six equations for the three perturbed metric functions $\nu_1,\Omega_1,\lambda_1$, and the two components $\xi^r,\xi^\theta$ of the gauge vector. All of these depend only on the coordinates $(r,\theta)$. 

Let us define a set of equations $\vec{\mathbb{E}}$ as follows
\begin{equation}
    \vec{\mathbb{E}}\equiv\vec{h}-\vec{\tilde{h}}-\Lie{\xi}\vec{g} =0 \,.
\label{eq:caleqE}
\end{equation}
We list all nontrivial components of $\vec{\mathbb{E}}$ in Appendix~\ref{app:csce}. Most of them contain a combination of the derivatives of the Debye potential, the linearized WLP metric functions, and the components of the gauge vector. The component $\vec{\mathbb{E}}_{r\theta}$, however, contains neither $\nu_1$, $\lambda_1$ nor $\Omega_1$, and reduces to 
\begin{align}
\frac{\Sigma}{2\Delta}\left(z+\bar{z}\right) &=\frac{\p}{\p r}\,\xi^\theta+\frac{1}{\Delta}\,\frac{\p}{\p\theta}\,\xi^r\,. \label{eq:cal1}
\end{align}
Furthermore, the linear combination
\begin{equation}
    -\Upsilon\, \mathbb{E}_{tt} - 4\ka \mass r \, \mathbb{E}_{t\phi} + \frac{\Sigma_m^-}{\sin^2\theta}\, \mathbb{E}_{\phi\phi}\,,
\end{equation}
where $\Upsilon$ is defined by Eq.~(\ref{eq:UpsilonDef}), yields a purely algebraic constraint on the gauge vector,
\begin{equation}
  \frac{\Delta'}{\Delta}\, \xi^r + 2\xi^\theta\cot\theta\, =\frac{\Sigma}{2}\left[\frac{1}{\Delta}\left(x+\bar{x}\right)+\left(y+\bar{y}\right)\right] ,\label{eq:cal2}
\end{equation}
which we can use to express Eq.~(\ref{eq:cal1}) in terms of a single component of the gauge vector. Moreover, the difference $\mathbb{E}_{\theta\theta}-\Delta\mathbb{E}_{rr}$, together with the Eq.~(\ref{eq:cal2}), yields
\begin{equation}
   \frac{\Sigma}{2} (y+\bar{y}) = \frac{\p}{\p r}\xi^r-\sin\theta \frac{\p}{\p\theta}\frac{\xi^\theta}{\sin\theta}\,.\label{eq:cal3}
\end{equation}
By explicitly solving Eqs.~(\ref{eq:cal1})\,--\,(\ref{eq:cal3}), we have found the components of the gauge vector in terms of the derivatives of the Debye potential, and from the remaining components of Eq.~(\ref{eq:caleqE}) (see Appendix~\ref{app:csce}) we have also expressed the perturbations of the metric functions. Altogether, the results read
\newcommand{\DD}{\tilde{\Delta}}
\newcommand{\Dx}{\hat{\Delta}}
\begin{widetext}
\begin{align}
    \xi^r &= \frac{1}{2\Sigma\DeltaPlus \sin\theta}
    \left[a\left(\Dx \Xi_{\mathrm{i},\theta}+\sin\theta\cos\theta\DD \Xi_{\mathrm{i},r}\right)
    +\left(\DD\cos\theta\, \Xi_{\mathrm{r},\theta}-\sin\theta\left((\mass-r)\DD-\ka^2\DeltaPlus \right)\Xi_{\mathrm{r},r}\right)\right]\,, \label{eq:xir}\\
    \xi^\theta &= \frac{1}{2\Sigma\DeltaPlus \sin^2\theta}
    \left[\phantom{a}\left(\Dx \Xi_{\mathrm{r},\theta}+\sin\theta\cos\theta\DD \Xi_{\mathrm{r},r}\right)
    +\ka\cos\theta\left(\Dx_{,r}\Xi_{\mathrm{i},\theta}+\tan\theta\Dx\Xi_{\mathrm{i},r}\right)\right]\,,\\
    \nu_1 &= \frac{1}{4}\frac{\Sigma}{\Sigma_m^-}\left[(x+\bar{x})+2ia\sin\theta\left(z-\bar{z}\right)-a^2\sin^2\theta\left(y+\bar{y}\right)\right]
    +\frac{\mass}{\Sigma_m^{-}\Sigma}\left(-\sigma \xi^r+r a^2\sin 2\theta \,\xi^\theta\right), \label{eq:nu1}\\
    \lambda_1 &=\nu_1-\frac{1}{4}\Sigma\left(y+\bar{y}\right)-\xi^\theta_{\ ,\theta}-\frac{1}{ 2\Sigma} \left(\xi^r\Sigma_{,r}+\xi^\theta\Sigma_{,\theta}\right)\,, \\
    \Omega_1 &= \frac{\Sigma}{2\Sigma_m^-}\,\sin^2\theta\Biggl[a(x+\bar{x})\,\frac{\Sigma_m^+}{\Sigma_m^-}-a(y+\bar{y})\Delta    
    +i\frac{(z-\bar{z})}{\sin\theta}\frac{\Sigma}{\Sigma_m^-}\left(\Delta+\ka^2\sin^2\theta\right)\Biggr] 
    -\frac{2 \ka\mass\sin^2\theta\left(2\sigma+\Sigma_m^-\right)}{\left(\Sigma_m^-\right)^2}\,\xi^r\,, \label{eq:Omega1}
\end{align}
\end{widetext}
where we introduced several new functions\footnote{$\Sigma_m^\pm$ is defined by Eq. (\ref{eq:SigmapmDef}) and $\DeltaPlus$ is explicitly worked out product of functions defined in Eq.~(\ref{eq:RpRm}) in BL coordinates.}
\begin{align}
    \DD &= \mass(r^2+\ka^2)-2\ka^2r \,, \\
    \Dx &= r^2-\mass r\sin^2\theta-\ka^2\cos^2\theta\,,\\
    \DeltaPlus  &= \Delta + \beta^2\sin^2\theta \,, \label{eq:Delta+} \\
    \sigma &= r^2-\ka^2\cos^2\theta\,,
\end{align}
and split the Debye potential multiplied by $\sin^2\theta$ into its real and imaginary parts
\begin{align}
    \chi_{[-4,0]} &= \frac{1}{\sin^2\theta}\left(\Xi_{\mathrm{r}}+i\,\Xi_{\mathrm{i}}\right). \label{eq:ChiToXiDef}
\end{align}

By expressing the perturbed metric functions in Weyl coordinates using the transformation~(\ref{eq:toW}), we have verified that the metric functions reconstructed from any stationary and axially symmetric solution of the Debye equation~(\ref{eq:GHPDebMin}) satisfy the linearized Einstein equations (\ref{eq:linNu})\,--\,(\ref{eq:linGammar}).


\section{Pure type D perturbations}
\label{sec:typeDpert}

Suppose we have a solution $\dot{\psi}_0$ or $\dot{\psi}_4$ of the Teukolsky equation describing a chosen physical system at hand. The corresponding Debye potential in one of the radiation gauges is then obtained by solving Eqs. (\ref{eq:psi0fromDebMin}), (\ref{eq:psi4fromDebMin}), and Eqs. (\ref{eq:psi0fromDebMax}), (\ref{eq:psi4fromDebMax}), i.e. by four-fold integration. In general, the Debye potential consists of two contributions
\begin{equation}
    \bar{\chi} = \bar{\chi}^\text{P} + \bar{\chi}^\text{H} \,,
\end{equation}
where $\bar{\chi}^\text{P}$ is a particular solution fixed by the values of $\dot{\psi}_0$ or $\dot{\psi}_4$, while $\bar{\chi}^\text{H}$ is a homogeneous solution $\dot{\psi}_0[\bar{\chi}^\text{H}] = 0$, and $\dot{\psi}_4[\bar{\chi}^\text{H}] = 0$. As Wald~\cite{wald1973:perturbationsKerr} showed, the homogeneous solutions lead, up to pure-gauge modes, to perturbations of the Kerr background that remain within the type D vacuum class. In this section, we analyze in detail the physical content of the homogeneous part of the Debye potential~$\bar{\chi}^\text{H}$ which may have to be added to $\bar{\chi}^\text{P}$ in order to match regularity condition on the axis and the required asymptotic behavior.

The complete set of four independent solutions of the Debye equation (\ref{eq:GHPDebMin}) compatible with $\dot{\psi}_0 = 0$ ($\dot{\psi}_4 = 0$ is satisfied automatically \cite{wald1973:perturbationsKerr}) reads
\begin{align}
    \bar{\chi}_{[0,-4]}^\text{H} &=\phantom{+} \Delta(r) \tilde{\mathrm{Q}}^2_{-2}\left(+\frac{1}{2}\cst{-2,1}\hat{\mathrm{Q}}_1^2-\frac{1}{3}\cst{-2,2}\hat{\mathrm{Q}}_{-2}^2\right)\nonumber\\
    &\phantom{=}+\Delta(r) \tilde{\mathrm{Q}}^2_{-1}\left(-\frac{1}{2}\cst{-1,1}\hat{\mathrm{Q}}_0^2-2\cst{-1,2}\hat{\mathrm{Q}}_{-1}^2\right)\nonumber\\
    &\phantom{=}+\Delta(r) \tilde{\mathrm{Q}}^2_{0\phantom{-}}\left(-\frac{1}{2}\cst{0,1}\hat{\mathrm{Q}}_0^2-2\cst{0,2}\hat{\mathrm{Q}}_{-1}^2\right)\nonumber\\
    &\phantom{=}+ \Delta(r) \tilde{\mathrm{Q}}^2_{1\phantom{-}}\left(+\frac{1}{2}\cst{1,1}\hat{\mathrm{Q}}_1^2-\frac{1}{3}\cst{1,2}\hat{\mathrm{Q}}_{-2}^2\right),
    \label{eq:HomogSol}
\end{align}
where we introduced the tilde and hat shorthand as follows:
\begin{align}
    \tilde{\mathrm{Q}}_l^s &= \mathrm{Q}^s_{l}\!\left(\frac{\Delta'(r)}{2\beta}\right), &
    \hat{\mathrm{Q}}_l^s &=\mathrm{Q}_l^s(\cos\theta)
\end{align}
with $\mathrm{Q}_l^s$ being associated Legendre functions of the second kind, explicitly:
\begin{align}
    \mathrm{Q}^2_{-2}(v) &= -\frac{v(v^2 - 3)}{v^2 - 1} \,, \\
    \mathrm{Q}^2_{-1}(v) &= \frac{v^2 + 1}{v^2 - 1} \,, \\
    \mathrm{Q}^2_{0} (v) &= \frac{2v}{v^2 - 1} \,, \\
    \mathrm{Q}^2_1 (v) &= \frac{2}{v^2 - 1} \,.
\end{align}
Evidently, the perturbations towards general type D spacetime are generated by the irregular $\bar{\chi}^\text{H}_{[0, -4]}$, which in this gauge is singular on the symmetry axis $\cos\theta = \pm 1$. This is consistent with the result of van de Meent \cite{vandemeent2017:massangular}, who showed that any metric perturbation reconstructed from a regular harmonic Debye potential carries vanishing mass and angular momentum perturbations.

The full solution (\ref{eq:HomogSol}) contains eight undetermined complex constants, i.e., 16 real free parameters. To identify the physical meaning of all of them, we consider the most general vacuum type D solution, the Pleba\'{n}ski\,--\,Demia\'{n}ski (PD) class~\cite{Plebanski1976}, in its newest $A^+$ form~\cite{ovcharenko_revisiting_2025}), linearized around the Kerr black hole with respect to its various parameters. Having the metric perturbations both (a) from the linearization of full PD solution and (b) reconstructed from the Debye potential via Eq.~(\ref{eq:sum}), we determine the gauge vector relating the two gauges and thereby relate the physical parameters of the type D spacetimes to the parameters $\mathfrak{A,B,C,D}$ of (\ref{eq:HomogSol}).

A remark on the Podolsk\'y\,--\,Vr\'atn\'y form \cite{podolsky_new_2021} of the PD metric: when linearizing this form, we would have to set $\Im(\cst{-2,2})=0$ in order to find the gauge vector, which would malevolently switch off some of the physical parameters. In this way, the $A^+$ form is more general. 

The $A^+$ form of PD metric reads as follows
\begin{align}
    \vec{\d} s_\mathrm{PD}^2 &= -\frac{1}{\Omega^2}\Bigl[\frac{\Delta_r}{\Sigma_\mathrm{PD}}\left(A\vec{\d} t-B\vec{\d}\phi\right)^2 +\frac{\Delta_x}{\Sigma_\mathrm{PD}}\left(C\vec{\d} t+D\vec{\d}\phi\right)^2\nonumber\\
    &\qquad+C_{\!f}\,\Sigma_\mathrm{PD}\left(\frac{\vec{\d} r^2}{\Delta_r}+\frac{\vec{\d} x^2}{\Delta_x}\right)\Bigr],
\end{align}
where 
\begin{align}
    A&=1+\alpha^2(l^2-a^2)x^2\,, &
    B&=(1+x^2)a+2lx\,,\\
    C&=a(1+\alpha^2r^2)+2\alpha lr \,,&
    D&=l^2-a^2+r^2\,,
\end{align}
and
\begin{align}
    \Omega &= 1-\alpha rx\,,\\
    \Delta_x &= (1-x^2)\left[(1-\alpha m x)^2-\alpha^2x^2(m^2+l^2-a^2)\right],\\
    \Delta_r &= (1-\alpha^2r^2)\left[(r-m)^2-(m^2+l^2-a^2)\right],\\
    \Sigma_\mathrm{PD} &= AD+BC\,,
\end{align}
where $m$ is the mass parameter, $a$ the rotation parameter, $\alpha$ the acceleration parameter and $l$ the NUT parameter. Finally $C_{\!f}$ is a real constant that we fix $C_{\!f}=1/(1+\alpha^2\ka^2)$ according to \cite{ovcharenko_revisiting_2025}.

The vacuum PD class is a four-parameter family of solutions, whereas the homogeneous solution~(\ref{eq:HomogSol}) carries 16 real parameters. However, not all additional freedom is pure gauge. In particular, the following coordinate transformation
\begin{align}
    \vec{\d} t &\rightarrow K_t\, \vec{\d} t+L\,\vec{\d}\phi\,, &
    \vec{\d} \phi \rightarrow K_\phi\,\vec{\d}\phi+\gamma\,\vec{\d} t\,, \label{eq:GlobalGauge}
\end{align} 
where $K_t$, $L$, $K_\phi$, and $\gamma$ are constants, preserves \emph{local} Einstein equations, but it changes the global or asymptotic structure of the spacetime. To fully identify the physical content of (\ref{eq:HomogSol}), we must therefore reintroduce this freedom explicitly. The individual parameters have the following meaning: $K_\phi$ controls the conical singularity along the axis, $K_t$ rescales the asymptotic time, and $\gamma$ sets the rotation of the asymptotic frame, while leaving the axis regular. Finally, $L$ uniformly twists the axis and, together with~$l$, controls the strength and distribution of the Misner string between the two half-axes; it is equivalent to the Manko--Ruiz parameter $C$ of Ref.~\cite{manko2005:Physicalinterpretation})\footnote{There is a common misconception that the singularity on the axis brought by NUT charge can be transformed to the other part of the axis. The local form of the metric is deceptive --- the singularity is present due to the gluing (the identification of points across the boundaries of $\phi$ coordinate). If we just perform coordinate transformation \emph{without} imposing new gluing conditions, we chang the spacetime. Inevitably, there must be two NUT charges whose ratio is controlled by $L$.}. In addition, we introduce a constant conformal rescaling
\begin{align}
    \vec{\d} s^2_\mathrm{PD}&\rightarrow \Omega_0^2\,\vec{\d} s^2_\mathrm{PD}\,,
    \label{eq:PDAgen}
\end{align}
on which we comment later.

Setting $x=\cos\theta$, we find the Kerr metric (\ref{eq:KerrMetric}) to be a special case of the $A^+$ form of PD metric for the background values
\begin{align}
    K_t&=-1,\,& 
    K_\phi&=1,\,& 
    L&=2\ka,\,&
    \gamma&=0,\,\nonumber\\
    l&=0,\, &\alpha&=0,\, & \Omega_0&=1\,.
    \label{eq:AplusKerr}
\end{align} 

Now, we calculate the linearized metric as follows
\begin{align}
    \vec{h}_{ab}(\dot{w}) &= \frac{\d}{\d\epsilon} \vec{g}_{ab}(w+\epsilon \dot{w})\Bigg|_{\epsilon=0}\,,
\end{align}
where $w$ stands for any single parameter of the set $\mathcal{W}=(m,a,\alpha,l,L,K_t,K_\phi,\gamma,\Omega_0)$ and the values of the other parameters are kept fixed at their background values (\ref{eq:AplusKerr}).

Our goal is now to solve for the gauge vector $\vec{\xi}^a$ which connects the metric perturbation $\vec{h}_{ab}[\chi]$ with the perturbation $\vec{h}_{ab}(\dot{w})$ as
\begin{align}
    \vec{h}_{ab}[\chi]+\nabla_{\!a}\vec{\xi}_b+\nabla_{\!b}\vec{\xi}_a-\sum_{\dot{w}\in \dot{\mathcal{W}}}\vec{h}_{ab}(\dot{w})&=0\,,
    \label{eq:genCal}
\end{align}
to identify 9 physical perturbations $\dot{\mathcal{W}}$ among the 16 real parameters of homogeneous solution, and to examine the gauge freedom to ensure that there is no physical perturbation left unrecognized\footnote{
The transformation (\ref{eq:GlobalGauge}), linearized around the background values~(\ref{eq:AplusKerr}), can be generated by
\begin{multline}
    \vec{\xi}_{\mathrm{glob}} = \left[ (2 \ka \dot{\gamma} - \dot{K}_t) t + (2\ka \dot{K}_\phi - \dot{L}) \phi \right] \vec{\p_t} + \\ 
    + \left[ \dot{\gamma} t + \dot{K}_\phi \phi \right] \vec{\p_\phi} \,. \nonumber
\end{multline}
Unlike the ansatz~(\ref{eq:GaugeVectorAnsatz}) employed in Sec.~\ref{sec:Calibration} and in the rest of the present Sec.~\ref{sec:typeDpert}, this generator depends explicitly on $t$ and $\phi$. Nevertheless, $\Lie{\xi_\mathrm{glob}}\vec{g}$ is again of the circular form, since~$\vec{\p_t}$ and~$\vec{\p_\phi}$ are Killing vectors and the components of~$\vec{\xi}_{\mathrm{glob}}$ are linear in $t$ and $\phi$ with constant coefficients. Up to the Killing vectors themselves, $\vec{\xi}_{\mathrm{glob}}$ is in fact the most general extension of~(\ref{eq:GaugeVectorAnsatz}) which respects symmetries of the spacetime. $\vec{\xi}_{\mathrm{glob}}$ does not, however, generate a proper gauge transformation: the terms proportional to $\phi$ are not single-valued once $\phi$ is periodically identified, and those proportional to $t$ do not fall off at infinity, which is precisely why the four constants $\dot{K}_t$, $\dot{L}$, $\dot{K}_\phi$ and $\dot{\gamma}$ carry physical information instead of being pure gauge.
}.

We consider the following ansatz for the gauge vector in NP basis:
\begin{align}
    \sqrt{2}\,\vec{\xi}^a & = 
    u^l \,\vec{n}^a
    +\frac{\Delta}{\Sigma}\,u^n\, \vec{l}^a
    -\frac{\sin\theta}{\rho}\,u^{\bar{m}}\, \vec{m}^a
    -\frac{\sin\theta}{\bar{\rho}}\,u^m\, \vec{\bar{m}}^a \,.
\end{align}
The pure gauge vector in BL coordinates is of the form $\vec{\xi}^a=\xi^r\vec{\p_r}^a+\xi^\theta\vec{\p_\theta}^a$, from which it follows that $u^l=-u^n$, while $u^{\bar{m}}$ is the complex conjugate of $u^m$. Thus, in what follows, we list only components $u^l$ and $u^m$.

Since the general solution of Eq.~(\ref{eq:genCal}) is lengthy, we leave its full form to the accompanying \emph{Wolfram Mathematica}$^{\copyright}$ notebook \cite{10.5281/zenodo.21699521}. In the following, we set the free parameters in such a way that components of the gauge vector are as simple as possible:

\begin{itemize}
\item
Mass perturbation $\mass+\epsilon \dot{m}$ 
\begin{fleqn}[2em]
\begin{equation}
\begin{split}
    \eqlhs{(u^l,u^m)} &= \dot{m}(\sigma/\Delta,\,2r\cot\theta/\sin\theta)\,,\\
    \eqlhs{\cst{0,2}}  &= -\dot{m}/(2\beta)\,,\\
    \eqlhs{\cst{1,2}}  &= -i\dot{m}\ka/\beta^2\,.
\end{split}
\end{equation}
\end{fleqn}
Different choice of free parameters yields the metric reconstructed from the Debye potential directly in the gauge obtained by linearization of the Kerr metric. This choice, however, is not suitable for the Schwarzschild limit
\begin{fleqn}[2em]
\begin{equation}
\begin{split}
    \eqlhs{(u^l,u^m)} &= (0,\,0)\,,\\
    \eqlhs{\cst{-1,1}} &= -2i\dot{m}/\ka\,,\\
    \eqlhs{\cst{0,1}}  &= -2i\dot{m}\mass/(\ka\beta)\,,\\
    \eqlhs{\cst{1,2}}  &= -i\dot{m}\ka/\beta^2\,.
\end{split}
\end{equation}
\end{fleqn}
\item
Angular perturbation $\ka+\epsilon\dot{a}$
\begin{fleqn}[2em]
\begin{equation}
\begin{split}
    \eqlhs{(u^l,u^m)} &= \dot{a}\ka(1+\cos^2\theta)\left(r/\Delta,\, \cot\theta/\sin\theta\right),\\
    \eqlhs{\cst{0,1}}  &= 2i\dot{a}/\beta\,,\\
    \eqlhs{\cst{1,2}}  &= i\dot{a}\mass/\beta^2\,.
\end{split}
\end{equation}
\end{fleqn}
\item
Acceleration perturbation $\epsilon\dot{\alpha}$
\begin{fleqn}[2em]
\begin{equation}
\begin{split}
    \eqlhs{u^l}        &= -\dot{\alpha}(8r^2(r^2+\ka^2)+\ka^4)\cos\theta/(8\Delta)\,,\\
    \eqlhs{u^m}        &= \dot{\alpha}\ka^2r\cos4\theta/(8\sin^2\theta)\,,\\
    \eqlhs{\cst{-2,2}} &= 2\dot{\alpha}\beta\,,\\
    \eqlhs{\cst{-1,1}} &= -4\dot{\alpha}\mass\,,\\
    \eqlhs{\cst{0,1}}  &= -\dot{\alpha}(4\beta+\ka^2/(8\beta))\,,\\
    \eqlhs{\cst{1,2}}  &= \dot{\alpha}\mass(2+\ka^2/\beta^2)\,.
\end{split}
\end{equation}
\end{fleqn}
\item
NUT parameter $\epsilon \dot{l}$
\begin{fleqn}[2em]
\begin{equation}
\begin{split}
    \eqlhs{u^l}        &= -\dot{l}\ka r(r-2\mass)\cos\theta / (\mass\Delta)\,,\\
    \eqlhs{u^m}        &= -\dot{l}\ka (r-2\mass)\cot^2\theta / \mass\,,\\
    \eqlhs{\cst{-1,2}} &= -i\dot{l}/(2\mass)\,,\\
    \eqlhs{\cst{0,1}}  &= \dot{l}\ka/\beta\mass\,.
\end{split}
\end{equation}
\end{fleqn}
\item
Misner-string parameter $L+\epsilon \dot{L}$
\begin{fleqn}[2em]
\begin{equation}
\begin{split}
    \eqlhs{(u^l,u^m)} &= -\dot{L}\ka\left(r/\Delta,\,\cot\theta/\sin\theta\right),\\
    \eqlhs{\cst{0,1}}  &= -i\dot{L}/\beta\,.
\end{split}
\end{equation}
\end{fleqn}
\item
Conical deficit $K_\phi+\epsilon \dot{K}_\phi$
\begin{fleqn}[2em]
\begin{equation}
\begin{split}
    \eqlhs{u^l}        &= -\dot{K}_\phi r\ka^2(2\cos\theta-\sin^2\theta)/\Delta,\,\\
    \eqlhs{u^m}        &= \dot{K}_\phi (\sigma+(r^2+\ka^2)\cos\theta)/\sin^2\theta, \\
    \eqlhs{\cst{-2,1}} &= 2\dot{K}_\phi \beta/(3\mass)\,, \\
    \eqlhs{\cst{-1,1}} &= 2\dot{K}_\phi(-1+i\ka/\mass)\,, \\
    \eqlhs{\cst{-1,2}} &= i\dot{K}_\phi \ka/(2\mass)\,, \\
    \eqlhs{\cst{0,1}}  &= 2\dot{K}_\phi(-\mass+i \ka)/\beta\,, \\
    \eqlhs{\cst{0,2}}  &= -\dot{K}_\phi\ka^2 /(2\beta\mass)\,, \\
    \eqlhs{\cst{1,2}}  &= -i\dot{K}_\phi\ka/\mass \,.
\end{split}
\end{equation}
\end{fleqn}
\item
Conformal factor $\Omega_0+\epsilon\dot{\Omega}_0$
\begin{fleqn}[2em]
\begin{equation}
\begin{split}
    \eqlhs{u^l}        &= -2\dot{\Omega}_0\sigma(r-2\mass)/\Delta\,,\\
    \eqlhs{u^m}        &= -2\dot{\Omega}_0(\sigma-4\mass r)\cot\theta/\sin\theta\,,\\
    \eqlhs{\cst{-1,2}} &= \dot{\Omega}_0\,, \\
    \eqlhs{\cst{0,1}}  &= 4i\dot{\Omega}_0\ka/\beta\,, \\
    \eqlhs{\cst{0,2}}  &= -\dot{\Omega}_0\mass/\beta\,.
\end{split}
\end{equation}
\end{fleqn}
\item
Time dilation $K_t+\epsilon\dot{K}_t$
\begin{fleqn}[2em]
\begin{equation}
\begin{split}
    \eqlhs{u^l}        &= \dot{K}_t (2(r-3\mass)\sigma-\ka^2 r(3+\cos2\theta))/(2\Delta) \,,\\
    \eqlhs{u^m}        &= \dot{K}_t (\Delta+\sigma-4\mass r-2\ka^2)\cot\theta/\sin\theta\,,\\
    \eqlhs{\cst{-1,2}} &= -\dot{K}_t\,,\\
    \eqlhs{\cst{0,1}}  &= -4i\dot{K}_t\ka/\beta\,,\\
    \eqlhs{\cst{0,2}}  &= \dot{K}_t\mass/(2\beta)\,.
\end{split}
\end{equation}
\end{fleqn}
\item
Asymptotic-frame-rotation parameter $\epsilon\dot{\gamma}$
\begin{fleqn}[2em]
\begin{equation}
\begin{split}
    \eqlhs{u^l}        &= -\dot{\gamma}\,\ka r(r^2-\ka^2)(3+\cos2\theta)/(2\Delta)\,,\\
    \eqlhs{u^m}        &= -\dot{\gamma}\,\ka(r^2-\ka^2)(3+\cos2\theta)\cot\theta/(2\sin\theta)\,,\\
    \eqlhs{\cst{-2,2}} &= -2i\dot{\gamma}\beta\,,  \\
    \eqlhs{\cst{1,2}}  &= -2i\dot{\gamma}\mass\,.
\end{split}
\end{equation}
\end{fleqn}
\end{itemize}

Out of these 9 parameters only eight are linearly independent, as can be checked by solving
\begin{align}
    \nabla_{\!a}\vec{\xi}_b+\nabla_{\!b}\vec{\xi}_a-\sum_{\dot{w}\in \dot{\mathcal{W}}}\vec{h}_{ab}(\dot{w})&=0\,.
\end{align}
For example, we can gauge away $\dot{\Omega}_0$ by
\begin{equation}
\begin{split}
    (u^l,u^m) &= \dot{\Omega}_0(-r \Sigma/\Delta,\,0)\,,\\
    \dot{m} &= -\dot{\Omega}_0\mass\,,\\
    \dot{a} &= -\dot{\Omega}_0\ka\,,\\
    \dot{L} &= -2\dot{\Omega}_0\ka\,,
\end{split}
\end{equation}
but we believe that it may still be useful in some applications to introduce $\dot{\Omega}_0$.

Now, suppose we have a Debye potential $\bar{\chi}_{[0,-4]}$ which was obtained by fourfold integration of $\dot{\psi}_0$. In order to select the proper values of parameters of homogeneous solution, we expand the potential at radial infinity, compare with asymptotic behavior of homogeneous solution and use the gauge freedom to set to zero all the parameters that we do not wish to perturb. We consider the only admissible perturbations to be $\dot{m}$ and $\dot{a}$ --- and these are to be computed from the Abbott\,--\,Deser (AD) charges \cite{vandemeent2017:massangular,abbott_stability_1982} as charges associated with appropriate Killing vectors. Then, the contribution of the perturbation can be identified.
The full expression of the constants entering the homogeneous solution in terms of physical parameters is as follows
\begin{equation}
\begin{split}
    \cst{-2,1} &= \frac{2\dot{K}_\phi\beta}{3\mass}+i\,\cst{-2,1}^\mathrm{i}\,,\\
    \cst{-2,2} &= 2(\dot{\alpha}-i\dot{\gamma})\beta\,,\\
    \cst{-1,1} &= \cst{-1,1}^\mathrm{r}+i\cst{-1,1}^\mathrm{i}\,,\\
    \cst{-1,2} &= \dot{\Omega}_0-\dot{K}_t + i\left[ \frac{\beta\cst{0,2}^\mathrm{i}}{\mass}  - \ka \dot{\alpha} -\frac{\dot{l}}{2\mass} - \frac{\ka \cst{-1,1}^\mathrm{r}}{4\mass} \right] \,,\\
    \cst{0,1}  &= \cst{0,1}^\mathrm{r} +  i \frac{2\dot{a}-\dot{L}+4\ka(\dot{\Omega}_0-\dot{K}_t) + \mass\cst{-1,1}^\mathrm{i}}{\beta}\,,\\
    \cst{0,2}  &= \frac{\mass (\dot{K}_t - 2\dot{\Omega}_0) - \dot{\mass}}{2\beta} - \frac{\ka \cst{-1,1}^\mathrm{i}}{4\beta} + i\,\cst{0,2}^\mathrm{i}\,,\\
    \cst{1,1}  &= \cst{1,1}^\mathrm{r}+i\,\cst{1,1}^\mathrm{i}\,,\\
    \cst{1,2}  &= \cst{1,2}^\mathrm{r} - i\left[\frac{\ka\dot{K}_\phi}{\mass}+2\mass\dot{\gamma}+\frac{(\ka\dot{m}-\mass\dot{a})}{\beta^2} \right]\,,
\end{split}
\end{equation}
where we denoted the real and the imaginary part of constants by superscripts ${}^\mathrm{r,i}$.

\subsubsection{Mass and angular momentum perturbations}

As an example we work out the mass and angular perturbation in greater detail.
For the mass perturbation we get 
\begin{align}
    \nu_1 &= \frac{-\dot{m}}{2\Sigma}\Biggl( 2(r+m)+m^2 \frac{r(3+\cos2\theta)-2m\sin^2 \theta}{R_\mathrm{p}R_\mathrm{m}}\nonumber\\
    &\qquad\qquad+ m^2\frac{2a^2r\sin^2 2\theta}{R_\mathrm{p}R_\mathrm{m}\Sigma_m^-}\Biggr)\,,\\
    \Omega_1 &= \frac{2\ka}{\Sigma_m^-}\left(\nu_1\Sigma\sin^2\theta-\frac{\dot{m}\mass^2r\sin^2 2\theta}{2R_\mathrm{p}R_\mathrm{m}}\right),\\
    \lambda_1 &= -\frac{\dot{m}\mass\sin^2\theta}{\Sigma_m^-}\Biggl(1+\frac{2\mass^2P_2(\cos\theta)}{2R_\mathrm{p}R_\mathrm{m}}-\frac{\mass^2\beta^2\sin^2 2\theta}{2R^2_\mathrm{p}R^2_\mathrm{m}}\Biggr)\,,
\end{align}
where $r$ and $\theta$ is given by~(\ref{eq:toW}) with the unperturbed value of $m$. The reconstructed metric functions $\nu_1$, $\lambda_1$, and $\Omega_1$ are exactly those of the mass perturbation of the Kerr black hole in Weyl coordinates, i.e. linearization of (\ref{eq:Knu})\,--\,(\ref{eq:KOmega}) with respect to the black hole mass $m$. 
Note that these expressions are not the same when linearizing Kerr black hole metric in BL coordinates, because the transformation between Weyl and BL coordinates depends explicitly on the mass and spin of the black hole. Thus we have to apply a gauge transformation given by the linearization of the transformation relations~(\ref{eq:toW}) with the gauge vector
\begin{align}
    \vec{\xi}^a &= \frac{\d}{\d\epsilon}r (m+\epsilon \dot{m})\Bigg|_{\epsilon=0} \vec{\p_r}^a + \frac{\d}{\d\epsilon}\theta (m+\epsilon \dot{m})\Bigg|_{\epsilon=0} \vec{\p_\theta}^a \nonumber\\
    &=\dot{m}\left(1+\frac{m\Delta'\sin^2\theta}{2R_\mathrm{p}R_\mathrm{m}}\right)\vec{\p_r}^a+\dot{m}\frac{m\sin2\theta}{2R_\mathrm{p}R_\mathrm{m}}\vec{\p_\theta}^a\,.
\end{align}

For the angular momentum perturbation (here we linearize the Kerr metric (\ref{eq:KerrMetric}) itself\footnote{Due to the non-trivial background parameters which takes $A^+$ form (after introducing the new parameters, Eq.~(\ref{eq:PDAgen})) of PD metric to the Kerr metric, the linearizations of $\d s^2_{\mathrm{PD}}$ and $\d s^2$ (Kerr) in the parameter $a\rightarrow a+\epsilon\dot{a}$ does not coincide; for physical interpretation is more natural to linearize directly the Kerr metric.}, not the PD in $A^+$ form) we have
\begin{align}
    \nu_1 &=\frac{\dot{a}\ka\mass}{R_\mathrm{p}R_\mathrm{m}\Sigma}\left(2r P_2 -\mass\sin^2\theta+\frac{2rW_\theta\sin^2\theta}{\Sigma_m^-}\right),\\
    \Omega_1 &=\frac{2\ka \dot{a}\Sigma\sin^2\theta}{\Sigma_m^-}\,\nu_1-\frac{2\dot{a}\mass r\sin^2\theta}{R_\mathrm{p}R_\mathrm{m}}\left(1-\frac{W_\theta}{\Sigma_m^-}\right),\\
    \lambda_1 &=\frac{\dot{a}\ka\mass^2\sin^2 \theta}{2R_\mathrm{p}R_\mathrm{m}\Sigma_m^-}\left(4P_2-\frac{\beta^2 \sin^2 2\theta}{R_\mathrm{p}R_\mathrm{m}}\right),
\end{align}
with $P_2=P_2(\cos\theta)$ and $W_\theta=2\ka^2\cos^2\theta-\mass^2\sin^2\theta$. The gauge vector is as follows
\begin{align}
\vec{\xi}^a  &= \frac{\d}{\d\epsilon}r (\ka+\epsilon \dot{\ka})\Bigg|_{\epsilon=0} \vec{\p_r}^a + \frac{\d}{\d\epsilon}\theta (\ka+\epsilon \dot{\ka})\Bigg|_{\epsilon=0} \vec{\p_\theta}^a \nonumber\\
&= -\frac{\dot{a}\ka\sin\theta}{2R_\mathrm{p}R_\mathrm{m}}\left(\Delta'\sin\theta\,\vec{\p_r}^a+2\cos\theta\,\vec{\p_\theta}^a\right).
\end{align}

For the mass and angular perturbation we also calculated the AD charges \cite{abbott_stability_1982} using the notation of \cite{vandemeent2017:massangular}. Abbott \& Deser \cite{abbott_stability_1982} introduced a 2-form $\vec{F}_{\!ab}$ for a Killing vector $\vec{k}^j$ as follows
\begin{align}
    \vec{F}_{\!ab}[\vec{k}^j] &= \frac{1}{8\pi}\left(\vec{k}^j\nabla_{[a}\vec{\bar{h}}_{b]j}+\vec{\bar{h}}_{j[a\!}\nabla_{b]}\vec{k}^j-\vec{k}_{[a\!}\nabla^j\vec{\bar{h}}_{b]j}\right)\,,
\end{align}
where $\vec{\bar{h}}_{ab}=\vec{h}_{ab}-\nicefrac{1}{2}\, \vec{g}_{ab}\vec{h}_j^{\phantom{j}j}$. Then the quasi-local charges associated with a particular Killing vector $\vec{k}^a$ and perturbation $\vec{h}_{ab}$ can be calculated as integrals over a closed 2-surface $\mathcal{S}$
\begin{align}
    Q(\vec{h},\vec{k},\mathcal{S}) &= \oint_\mathcal{S}\vec{\star F}+i\vec{F}\,.
\end{align}

For the mass perturbation $\vec{h}^{(\dot{m})}$ we get
\begin{align}
    Q(\vec{h}^{(\dot{m})},\vec{\KVxi},\mathcal{S}_{tr}) &= \dot{m}\,, &
    Q(\vec{h}^{(\dot{m})},\vec{\eta},\mathcal{S}_{tr}) &= \dot{m}\ka\,,
\end{align}
where we can explicitly observe that the change of the mass parameter $m\rightarrow m+\epsilon\dot{m}$ induces not only the change in physical mass, but also a change in total angular momentum (since $\ka$ is kept fixed).

For the angular perturbation $\vec{h}^{(\dot{a})}$ the charges are
\begin{align}
    Q(\vec{h}^{(\dot{a})},\vec{\KVxi},\mathcal{S}_{tr}) &= 0\,, &
    Q(\vec{h}^{(\dot{a})},\vec{\eta},\mathcal{S}_{tr}) &= \dot{\ka}m\,.
\end{align}
The integration is performed over a topological 2-sphere $\mathcal{S}_{tr}$ of constant, but otherwise arbitrary, radius on a constant-time slice.

\subsubsection{Cosmic string}
\label{sec:CosmicString}
The full WLP equations (\ref{eq:psi})\,--\,(\ref{eq:lambdaz}), as well as their linearized counterparts (\ref{eq:linNu})\,--\,(\ref{eq:linGammar}), are invariant under constant shifts of $\lambda$ and $\Omega$ ($\lambda_1$ and $\Omega_1$, respectively); these constants are fixed not by the (local) Einstein equations but by global conditions. A natural global condition to impose is the regularity of the symmetry axis. If there are no sources located \emph{on} the axis, we assume the function $\nu$ to be finite and smooth on the axis, i.e., $\nu\sim\nu_{(0)}(z)+\nu_{(1)}(z)\varrho+\nu_{(2)}(z)\varrho^2/2+\mathcal{O}(\GHPrho^3)$. Substituting this expansion into Eq.~(\ref{eq:psi}), we found that $\Omega\sim \Omega_{(2)}(z)\varrho^2/2+\Omega_{(3)}(z)\varrho^3/6$ around the axis. In the limit to the axis, the elementary flatness condition $X^{,a}X_{,a}/(4X)\rightarrow 1$, where $X$ is the squared norm of the axial Killing vector (see Chapter 19 in \cite{stephani_exact_2003} for details), then reduces to the requirement known from the static case,
\begin{align}
    \lambda(\varrho=0,z) &=0\,,
    \label{eq:regaxis}
\end{align}
Since the Kerr background satisfies $\lambda_0=0$ on the axis, we thus impose $\lambda_1=0$ there.

A constant $\lambda_1 \neq 0$ represents a conical deficit. On the Kerr background, it can be introduced either by (a) rescaling the angular variable by $K_\varphi$ or by (b) shifting $\lambda\rightarrow \lambda-1/2\ln(1+\delta)$. These operations \emph{are not} equivalent, since the shift in $\lambda$ amounts to rescaling $\phi$ by $(1+\delta)$ \emph{combined} with the reciprocal rescaling of the background parameters, $m\rightarrow m/(1+\delta)$, $a\rightarrow a/(1+\delta)$.

The metric functions $\lambda_1$ and $\Omega_1$ reconstructed from the Debye potential generally acquire nonzero constant parts. To remove them, we look for a Debye potential that shifts these two constants without affecting the gravitational potential $\nu_1$,
\begin{align}
    \nu_1[\bar{\chi}^\mathrm{H}] &= 0\,, &
    \Omega_1[\bar{\chi}^\mathrm{H}] &= \mathrm{const}\,, &
    \lambda_1[\bar{\chi}^\mathrm{H}] &= \mathrm{const}\,.
\end{align}
The solution reads
\begin{align}
    \bar{\chi}^\mathrm{H}_{[0,-4]} &= -2\Delta^2\sin^2\theta
    \Biggl[+A_0\frac{z}{\GHPrho^4}\nonumber\\
    &+\frac{\beta+i\ka}{3\mass\beta}A_+\frac{\left((z+\beta)^2+\GHPrho^2\right)^\frac{3}{2}}{\GHPrho^4}\nonumber\\
    &+\frac{\beta-i\ka}{3\mass\beta}A_-\frac{\left((z-\beta)^2+\GHPrho^2\right)^\frac{3}{2}}{\GHPrho^4}\Biggr], \label{eq:DebyeCosmicString}
\end{align}
where $A_\pm$ are real constants, while $A_0$ can be taken purely imaginary. In terms of the constants of the homogeneous solution
(\ref{eq:HomogSol}),
\begin{align}
    \cst{-2,1}=\frac{4}{3}\cst{0,2} &= \frac{-2}{3\mass}\Bigl(\beta\,(A_++A_-)+i\ka\,(A_+-A_-)\Bigr)\,,\\
     \cst{-1,1}=-2\cst{1,2} &= \frac{-2}{\mass}\Bigl(i\ka\,(A_++A_-)+\beta\,(A_+-A_-)\Bigr)\,.\\
     \cst{0,1}& =i\Im(A_0)/\beta\,,
\end{align}
which leads to
\begin{align}
    \nu_1 &= 0\,, & \lambda_1 &= (A_++A_-)\,, & \Omega_1 &=-\Im(A_0)\,.
\end{align}
An appropriate choice of the constants $A_\pm$, $A_0$ therefore restores the regularity of the axis.

\section{Particular examples}
\label{sec:Examples}
In this section we present two Debye potentials which, unlike the homogeneous solutions of Sec.~\ref{sec:typeDpert}, take the spacetime out of the type-D class: a \textbf{static thin disk} around a Schwarzschild black hole, obtained by linearizing an explicitly known exact solution \cite{kotlarik_black_2022}, and a \textbf{rotating point particle} on the axis of the Kerr black hole~\cite{kofron2020}. In the accompanying \emph{Wolfram Mathematica} notebook~\cite{10.5281/zenodo.21699521} we give both potentials, check that the reconstructed metric satisfies the linearized Einstein equations, and verify analytically that the disk case reproduces the solution of Ref.~\cite{kotlarik_black_2022}.
 
\subsection{Static limit and an explicit disk example}
\label{sec:StaticLimit}

In the limit $a \rightarrow 0$, i.e., for the Schwarzschild background, the reconstruction formulas~(\ref{eq:nu1})\,--\,(\ref{eq:Omega1}) simplify considerably:
\begin{align}
     2\, \nu_1 \sin^2\theta &= -\frac{1}{2} \frac{\p^2\, \Xi_\text{r}}{\p r^2} - \label{eq:nu1forSchw}\\ 
     & \quad + \frac{1}{\DeltaPlus } \left[ (r - \mass)\frac{\p \, \Xi_\text{r}}{\p r}  + \cot\theta \frac{\p\, \Xi_\text{r}}{\p \theta} \right], \nonumber\\
    2\,\lambda_1 \frac{\DeltaPlus }{\mass} &= - (r - \mass)\frac{\p^2\, \Xi_\text{r}}{\p r^2} - \cot\theta \frac{\p^2 \, \Xi_\text{r}}{\p r \,\p\theta} \label{eq:lambda1forSchw} \\ 
    & \quad + \frac{(r - \mass)^2  + \mass^2\cos^2\theta}{\DeltaPlus } \frac{\p\, \Xi_\text{r}}{\p r}  \nonumber \\ 
    & \quad + \frac{2(r - \mass) \cot\theta}{\DeltaPlus } \frac{\p\, \Xi_\text{r}}{\p \theta} \,, \nonumber\\
    2 \,\Omega_1 \sin\theta &= \frac{1}{r - 2\mass} \left[ r \, \frac{\p^2 \,\Xi_\text{i}}{\p r\, \p \theta} - 2\, \frac{\p\, \Xi_\text{i}}{\p \theta} \right],
\end{align}
where $\DeltaPlus $ is defined in Eq.~(\ref{eq:Delta+}).

Note that perturbations of the Schwarzschild black hole naturally split into a purely static part, encoded in the real part of the Debye potential, and a stationary (rotational) part, encoded in its imaginary part. 

To illustrate the procedure, we construct the Debye potential that reproduces the linearized static thin-disk solution around the Schwarzschild black hole obtained in Ref.~\cite{kotlarik_black_2022}. Starting from the known explicit expression for~$\nu_1$, we solve Eq.~(\ref{eq:nu1forSchw}) for the Debye potential, following an approach inspired by Linet~\cite{Linet1977}: we recast the Debye equation~(\ref{eq:GHPDebMin}) as an axially symmetric Laplace equation and solve it using the Generalized Axially Symmetric Potential (GASP) theory.

First, we introduce a new function $\Phi$ such that
\begin{equation}
    \chi_{[-4, 0]}(r,\theta) = \frac{\Xi_\text{r}(r, \theta) + i\, \Xi_\text{i}(r, \theta)}{\sin^2 \theta} = \Phi(r, \theta) \Delta^2 \sin^2\theta \,. \label{eq:DebToPhi}
\end{equation}
In Weyl coordinates (\ref{eq:toW}), the Debye equation (\ref{eq:GHPDebMin}) then reduces to
\begin{equation}
    \Delta_2\Phi=\frac{\p^2 \Phi}{\p \varrho^2} + \frac{5}{\varrho}\frac{\p \Phi}{\p \varrho} + \frac{\p^2 \Phi}{\p z^2} = 0 \,,
\end{equation}
i.e., to the axially symmetric Laplace equation in 7 dimensions. Within GASP theory, the solution at a general point can be obtained from its values on the symmetry axis through the integral
\begin{equation}
    \Phi (\varrho,z) = \frac{8}{3\pi}\int_0^\pi \Phi\left(0,z+i\varrho\cos\alpha\right)\sin^4\alpha\,\d\alpha \,. \label{eq:gasp}
\end{equation}
Substituting Eq.~(\ref{eq:DebToPhi}) into Ea.~(\ref{eq:nu1forSchw}) and expanding about the symmetry axis, we obtain the simple relation:
\begin{equation}
    \nu_1(r, 0) = - 4 r (r - 2\mass) \Phi(r, 0) + \mathcal{O}(\theta) \,.
\end{equation}
Given $\nu_1$ on the axis, this relation determines $\Phi$ there; the quadrature~(\ref{eq:gasp}) then yields $\Phi$ at a general point $(r, \theta)$. Finally, differentiating according to Eq.~(\ref{eq:lambda1forSchw}) provides the perturbation of the second metric function~$\lambda_1$. As discussed in Sec.~\ref{sec:CosmicString}, this procedure may introduce a conical singularity on the symmetry axis, since it fixes $\lambda$ only up to a constant. In so, we subtract the homogeneous solution~(\ref{eq:DebyeCosmicString}), with suitably chosen constants, from the resulting Debye potential.

Carrying out these steps for the thin disk of Ref.~\cite{kotlarik_black_2022}, we arrive at the Debye potential
\begin{widetext}
    \begin{multline}
        \Xi_\text{r} = \frac{\mathcal{M}}{3\mass} \left\{ - 6 M (b + 2z) + \frac{2\mass + b}{(\mass + b)^2}R_\mathrm{m}^3 + \frac{2\mass - b}{(\mass - b)^2} R_\mathrm{p}^3 \right.\\
        \left. + \frac{2M\sqrt{\varrho^2 + (z + b)^2}}{(\mass^2 - b^2)^2} \left[ (z+b)(3b^3 - \mass^2 (5b + 2z)) - 2\mass^2\varrho^2 \right] \right\},
    \end{multline}
\end{widetext}
where $R_\mathrm{p}$ and $R_\mathrm{m}$ are defined in Eq.~(\ref{eq:RpRm}), $\mathcal{M}$ is the total mass of the disk, and $b$ is a parameter with the dimension of length. Substituting it into Eqs.~(\ref{eq:nu1forSchw}) and~(\ref{eq:lambda1forSchw}) indeed reproduces $\nu$, and the part of $\lambda$ linear in the disk mass, given by Eqs.~(9) and~(27) of Ref.~\cite{kotlarik_black_2022} for the disk labels $(m, n) = (0, 1)$ used therein. 

\subsection{Rotating point particle on the axis}
So far the only explicitly known Debye potential for gravitational perturbation of Kerr background represent a rotating point particle (with aligned axis of rotation) on the symmetry axis \cite{kofron2020}. 

The Debye potential takes a very compact form:
\begin{align}
    \chi_{[-4,0]} &= C \frac{X^{3/2}}{\sin^2\theta}\,, \label{eq:DebyePPChi}\\
    X &= \frac{1}{4}\left(\Delta'(r)\cos\theta-\Delta'(r_0)\right)^2+\Delta(r)\sin^2\theta\,,
    \label{eq:DebyePP}
\end{align}
where the complex constant $C$ encodes the mass and the angular momentum of the particle, and $r_0>r_\mathrm{p}$ denotes its position on the axis.

We leave the full explicit forms of the functions $\nu_1$, $\lambda_1$, and $\Omega_1$, reconstructed using Eqs.~(\ref{eq:nu1})\,--\,(\ref{eq:Omega1}), to the accompanying \emph{Wolfram Mathematica} notebook~\cite{10.5281/zenodo.21699521}, where we also verify that the reconstructed metric satisfies the linearized Einstein equations also in this rotating case.

For a nonrotating (Schwarzschild) background we obtain
\begin{align}
    \nu_1 &= \frac{3}{4}\frac{z_0^2 - m^2}{\sqrt{\varrho^2 + (z - z_0)^2}} \,,\\
    \lambda_1 &= - \frac{3}{4\sqrt{\varrho^2 + (z - z_0)^2}}\left[ (m + z_0)R_p + (m - z_0)R_m \right] \,, \\
    \Omega_1 &= -3 m \frac{R_p - R_m - 2z_0}{R_p + R_m - 2m}  \sqrt{\varrho^2 + (z - z_0)^2} \\
    & \quad - \frac{3(m^2 - z_0^2)}{2}\frac{R_p + R_m + 2m}{R_p + R_m - 2m}\frac{z - z_0}{\sqrt{\varrho^2 + (z - z_0)^2}}\,, \nonumber
\end{align}
where $z_0 = r_0 - m$ is the position of the \emph{rotating} point particle in Weyl coordinates, and the constant in Eq.~(\ref{eq:DebyePPChi}) has been chosen as $C = 1 + i$. In $\nu_1$ we readily recognize the expected gravitational potential of a point particle on the axis. 

As anticipated, the metric functions $\lambda_1$ and $\Omega_1$ do not vanish on the axis; on either side of the particle, they approach a \emph{constant} value
\begin{align}
    \lim_{\varrho \rightarrow 0^{+}} \lambda_1
        &= -\frac{3m}{2}\,\operatorname{sgn}(z - z_0) \,, \label{eq:lambda1PPAxis} \\
    \lim_{\varrho \rightarrow 0^{+}} \Omega_1
        &= \frac{3(m + z_0)^2}{2}\,\operatorname{sgn}(z - z_0) \,, \label{eq:Omega1PPaxis}
\end{align}
which signals the presence of a string holding the point particle at a fixed distance from the black hole. Adding the solution~(\ref{eq:DebyeCosmicString}), we can remove the string on the part of the axis either above or below the particle, but not globally, since both limits (\ref{eq:lambda1PPAxis}) and (\ref{eq:Omega1PPaxis}) change sign across the location of the particle.

Owing to the linearity of the problem, we can construct the perturbation of the Kerr black hole not only by a single point particle, but by an arbitrary line distribution of matter along the symmetry axis. Such a field can then be used to generate a disk around the black hole, much as in Refs.~\cite{kotlarik_black_2022,vieira2020,kofron2023}; there, however, the results are exact but static, whereas the construction outlined here would yield a rotating disk, albeit only to the linear perturbation order. We leave a more detailed investigation to future work.

\section{Conclusions}
\label{sec:Conclusions}
We have found an explicit gauge transformation that connects two approaches to stationary and axially symmetric perturbations of the Kerr black hole: direct circular metric perturbations and perturbations within the NP/GHP formalism. Specifically, we have shown how to reconstruct the metric perturbation in a circular gauge --- the WLP form --- in terms of the Debye potential. On the Kerr background, both the Debye equation, (\ref{eq:GHPDebMax}) or (\ref{eq:GHPDebMin}), and the TME simplify to the 7-dimensional Laplace equation in cylindrical coordinates~\cite{Linet1977}. Instead of solving the linearized field equations (\ref{eq:linNu})\,--\,(\ref{eq:linGammar}), a coupled system for the metric perturbation, we can therefore obtain the WLP metric functions from a solution of a single linear Laplace equation.

We then discussed several special cases of the axially symmetric and stationary Debye potential. First, we considered the complete set of homogeneous solutions, i.e., those with $\dot{\psi}_0[\bar{\chi}^\text{H}] = 0$, and $\dot{\psi}_4[\bar{\chi}^\text{H}] = 0$, which correspond to pure type D perturbations of the Kerr black hole. By fully exploiting the gauge freedom, we linked coefficients of the general solution to a 9-parameter physical family: the Pleba\'{n}ski\,--\,Demia\'{n}ski class in the $A^{+}$ form of Ovcharenko, Podolsk\'y, and Astorino~\cite{ovcharenko_revisiting_2025}. In particular, we identified the Debye potential leading to the mass and angular momentum perturbations of the Kerr black hole. Next, we obtained the Debye potential leading to constant $\lambda_1$ and $\Omega_1$, i.e. describing a cosmic string on the Kerr background. 

Finally, we present two particular examples of the Debye potential corresponding perturbations that are not of type D. In the limit of the Schwarzschild background, we constructed an explicit Debye potential giving the perturbation of the Schwarzschild black hole by a thin disk, namely the linearized solution of Ref.~\cite{kotlarik_black_2022}. In the rotating case, we considered a point particle on the symmetry axis and reconstructed the corresponding WLP metric functions in closed form.

In future work, we plan to generalize Linet's approach \cite{Linet1977,Linet1979} by explicitly constructing perturbations of physical interest, such as a rotating ring or a thin disk encircling the Kerr black hole.

\begin{acknowledgments}
    The authors acknowledge support from the INTER-COST LUC25003 project of the INTER-EXCELLENCE II programme of the Czech Ministry of Education, Youth and Sports and the contribution of the COST Action CA23130. PK is grateful for the support by JSPS Postdoctoral Fellowships for Research in Japan (P25748) and the JSPS KAKENHI grant 25KF0237.

    The calculations were performed using \emph{Wolfram Mathematica}$^{\copyright}$ and the \emph{xAct} package \cite{xact} in the version 1.3.0.
\end{acknowledgments}

\onecolumngrid

\appendix

\section{NP quantities of the Kerr black hole}
\label{app:Kerr}

The nonzero NP spin coefficients corresponding to the tetrad (\ref{eq:NPtetrad}) and the only non-vanishing projection of the Weyl tensor are
\begin{equation}
\begin{aligned}
\pi &= \frac{i}{\sqrt{2}}\,\frac{\ka \sin\theta}{\Krho^2}\,, &
\mu &= \frac{-1}{\sqrt{2}}\,\frac{\Delta}{\Sigma\Krho}\,, &
\tau &= \frac{-i}{\sqrt{2}}\,\frac{\ka \sin\theta}{\Sigma}\,, &
\GHPrho &= \frac{-1}{\sqrt{2}}\,\frac{1}{\Krho}\,,\\
\gamma &=\mu +\frac{1}{\sqrt{2}}\,\frac{r-\mass}{\Sigma}\,, &
\beta &= \frac{1}{\sqrt{2}}\,\frac{\cot \theta}{\Krhocc}\,, &
\alpha &= \pi-\bar{\beta}\,, & \psi_2 & = -\frac{\mass}{\Krho^3}\,.
\end{aligned}\label{eq:spinc}
\end{equation}

\section{Coordinate form of \texorpdfstring{$x,y,z$}{x, y, z} scalars}
\label{sec:xyz}
The scalars $x,y,z$ which are derived from the Debye potentials acquire the following coordinate form
\begin{align}
    x &= \frac{1}{\bar{\rho}^2\sin^2\theta}\left(
    \frac{-i\ka+r\cos\theta}{\rho\sin\theta}\frac{\p }{\p\theta}+\frac{\Delta_{,r}}{2}\,\frac{\p}{\p r}- \frac{\Delta}{2}\frac{\p^2}{\p r^2} \right) \chi_{[-4,0]} \,\sin^2\theta\,,\\
    y &= \frac{1}{\rho^2}\left(-\frac{1}{\bar{\rho}}\frac{\p}{\p r}+\frac{1}{2}\frac{\p^2}{\p r^2}\right)\bar{\chi}_{[0,-4]}\,, \\
    z &= \frac{1}{\rho^3\bar{\rho}\sin^2\theta}\left(-r\frac{\p}{\p\theta}-\frac{\Sigma_{,\theta}}{2}\frac{\p}{\p r}+\frac{\Sigma}{2}\frac{\p^2}{\p r\p\theta}\right)  \bar{\chi}_{[0,-4]} \,\sin^2\theta\,,
\end{align}
where the operators in parentheses act on everything to the right of them.

\section{The complete set of gauge equations}
\label{app:csce}
In this Appendix, we list the full set of gauge equations~(\ref{eq:caleqE}) expressed in BL coordinates:
\begin{align}
    \mathbb{E}_{tt}&:&
    \frac{1}{2}\left((x+\bar{x})+2i\ka\sin\theta\,(z-\bar{z})-\ka^2\sin^2\theta\,(y+\bar{y}) \right)&=
    \frac{2\mass\sigma}{\Sigma^2}\,\xi^r
    -\frac{2\ka^2 \mass r\sin2\theta}{\Sigma^2}\,\xi^\theta
    +2\frac{\Sigma_m^-}{\Sigma}\,\nu_1 \,,\\
    \mathbb{E}_{rr} &:&
    \frac{\Sigma^2}{2\Delta^2}\,(x+\bar{x}) &=-\frac{1}{\Delta}\left(\xi^\theta\Sigma_{,\theta}+\xi^r\Sigma_{,r}\right)+\frac{\Sigma}{\Delta^2}\left(\xi^r\Delta_{,r}-2\Delta\xi^r_{\,,r}\right)+2\frac{\Sigma}{\Delta}\left(\nu_1-\lambda_1\right),\\
    \mathbb{E}_{r\theta} &:&
    \frac{\Sigma}{2\Delta}\,(z+\bar{z}) &=
    \frac{1}{\Delta}\,\xi^r_{\ ,\theta}+\xi^\theta_{\ ,r}\,,\\
    \mathbb{E}_{\theta\theta} &:&
    \frac{1}{2}\Sigma\, (y+\bar{y}) &= -\frac{1}{\Sigma}\left(\xi^\theta\Sigma_{,\theta}+\xi^r\Sigma_{,r}\right)-2\xi^\theta_{\ ,\theta}+2\nu_1-2\lambda_1\,,
\end{align}
and the two last equations, namely $\mathbb{E}_{t\phi}$, 
\begin{multline}
    \frac{1}{2}\Biggl[\ka\Bigl(-(x+\bar{x})+(r^2+\ka^2)(y+\bar{y})\Bigr)\sin\theta-i(\sigma+2\ka^2)(z-\bar{z})\Biggr] \sin\theta\\
=-\frac{2\ka \mass r^2\sin^4\theta}{\Sigma^2}\left[\xi^r\left(\frac{\Sigma}{r\sin^2\theta}\right)_{\!\!,r}+\xi^\theta\left(\frac{\Sigma}{r\sin^2\theta}\right)_{\!\!,\theta}\right]+\frac{4\ka \mass r\sin^2\theta}{\Sigma}\,\nu_1-\frac{\Sigma_m^-}{\Sigma}\,\Omega_1\,,
\end{multline}
and, finally, $\mathbb{E}_{\phi\phi}$,
\begin{multline}
    \frac{1}{2}\left(\ka^2\sin^2\theta\,(x+\bar{x})+2i\ka(\ka^2+r^2)\sin\theta\,(z-\bar{z})-(\ka^2+r^2)^2(y+\bar{y}) \right)\\
    =\frac{\Upsilon^2\sin^2\theta}{\Sigma^2}\left[\xi^r\left(\frac{\Sigma}{\Upsilon\sin^2\theta}\right)_{\!\!,r}+\xi^\theta\left(\frac{\Sigma}{\Upsilon\sin^2\theta}\right)_{\!\!,\theta}\right]+2\frac{4\ka^2\mass^2r^2\sin^2\theta+\Delta\Sigma^2}{\Sigma_m^-\Sigma}\,\nu_1-\frac{4\ka \mass r}{\Sigma}\,\Omega_1\,.
\end{multline}
\twocolumngrid

\section{Killing spinors, vectors and tensors of the Kerr background}
\label{app:Killing}
The Kerr metric admits a valence two Killing spinor \cite{Andersson2015}, a solution to the equation
\begin{equation}
    \nabla_{A'(A}\,\kappa_{BC)} = 0\,. 
    \label{eq:spinorEq}
\end{equation}
In adapted spinor dyad $o_A$, $\iota_A$, the Killing spinor takes the following simple form:
\begin{align}
    \kappa_{AB}=-2\kappa_1 o_{(A}\iota_{B)} \,.
\end{align}
Then, the four independent projections of the Eq.~(\ref{eq:spinorEq}) become equations for the $\kappa_1$ component and read as follows:
\begin{align}
\thorn\kappa_1 &= -\GHPrho\kappa_1 \,, & \eth\kappa_1 &=-\tau\kappa_1 \,, \\
\thorn'\kappa_1 &= -\GHPrho'\kappa_1 \,, & \eth'\kappa_1 &=-\tau'\kappa_1 \,.
\end{align}
In the Boyer\,--\,Lindquist coordinates, the solution of these equations is
\begin{equation}
\kappa_1 = \rho\,,  
\end{equation}
where $\rho=r-ia\cos\theta$ as defined by Eq.~(\ref{eq:Krho}).

For the simplification of GHP expressions we need to use some identities valid only on type D background in an adapted tetrad. Since these identities are scattered across literature \cite{edgar_petrov_2009,price_developmenst_2015} we will provide a short sketch of the derivation. The simplified versions of Ricci and Bianchi identities are not explicitly written out.

A Killing vector 
\begin{align}
    \vec{\tilde{\xi}_{(t)}}^a &= \frac{1}{3}\nabla^{A'B}\kappa^A_{\phantom{A}B}\\
    &=
    \kappa_1\left(
    \rho'\, \vec{l}^a
    -\rho\, \vec{n}^a
    -\tau'\, \vec{m}^a
    +\tau\,\vec{\bar{m}}^a\right)
\end{align}
can be in principle complex.

A two form $\vec{\hat{t}}_{ab}$ fulfiling equation \cite{frolov2018,penrose_naked_1973,frolov_black_2017}
\begin{align}
    \nabla_{\!c}\,\vec{\hat{t}}_{ab} &= \nabla_{[c}\,\vec{\hat{t}}_{ab]}+\vec{g}_{c[a}\vec{\varsigma}[\vec{\hat{t}}]_{b]}\,,\\
    \vec{\varsigma}[\vec{\hat{t}}]_{b} &=  \frac{1}{3}\nabla_{\!j}\,\vec{\hat{t}}^j_{\phantom{j}a}
\end{align}
is a rank 2 conformal Killing\,--\,Yano (CKY) tensor and $\vec{\varsigma}[\vec{t}]_b$ is either Killing vector or zero \cite{jezierski_conformal_2006}.

The existence of Killing spinor $\kappa_{AB}$ gives rise to the  CKY tensor $\vec{\hat{h}}$ as follows
\begin{align}
\vec{\hat{h}}_{ab}&=-\frac{1}{2}\left(\kappa_{AB}\bar{\epsilon}_{A'B'}+\epsilon_{AB}\bar{\kappa}_{A'B'}\right)\,,\\
&=
(\kappa_1+\bar{\kappa}_{1'})\vec{l}_{[a}\vec{n}_{b]}
-(\kappa_1-\bar{\kappa}_{1'})\vec{m}_{[a}\vec{\bar{m}}_{b]}\,.
\end{align}
According to \cite{krtous_killing-yano_2007,frolov_black_2017} for non-accelerated type D spacetimes this CKY should be a principal KY tensor --- a closed CKY tensor, so that we have
\begin{align}
    \nabla_{\!c}\,\vec{\hat{h}}_{ab} &= \vec{g}_{c[a}\vec{\KVxi}_{b]}\,, &
    \nabla_{[c}\,\vec{\hat{h}}_{ab]} &=0\,,
    \label{eq:PKY}
\end{align}
with
\begin{align}
    \vec{\KVxi}^a &= \vec{\varsigma}[\vec{\hat{h}]}^a=\frac{1}{2}\left(\vec{\tilde{\xi}_{(t)}}^a+\vec{\bar{\tilde{\xi}}_{(t)}}^a\right)=\vec{\p_t}\,.
\end{align}
being a non-trivial Killing vector.

The Eq.~(\ref{eq:PKY}) lead to new identities (which are easily verifiable in coordinates, this provides their geometrical origin) among the GHP scalars as follows
\begin{align}
    \kappa_1\GHPrho -\bar{\kappa}_{1'}\,\bar{\GHPrho} &=0\,, & \kappa_1\tau+\bar{\kappa}_{1'}\,\bar{\tau}' &=0\,, \\
    \kappa_1\GHPrho' -\bar{\kappa}_{1'}\,\bar{\GHPrho}' &=0\,, & \kappa_1\tau'+\bar{\kappa}_{1'}\,\bar{\tau} &=0\,,
\end{align}
and these identities automatically makes even the Killing vector $\vec{\tilde{\xi}_{(t)}}$ real.

A KY tensor can be constructed from $\vec{h}_{ab}$ using Hodge dualisation. 
Thus we have 
\begin{align}
    \vec{\hat{k}}_{ab} &= (\vec{\star \hat{h}})_{ab} = \frac{1}{2}\vec{\epsilon}_{abcd}\vec{\hat{h}}^{cd}\,, \\
    &=-\frac{i}{2}\left(\kappa_{AB}\bar{\epsilon}_{A'B'}-\epsilon_{AB}\bar{\kappa}_{A'B'}\right)\,,\\
&=
i\left[(\kappa_1-\bar{\kappa}_{1'})\vec{l}_{[a}\vec{n}_{b]}
-(\kappa_1+\bar{\kappa}_{1'})\vec{m}_{[a}\vec{\bar{m}}_{b]}\right] 
\end{align}
due to the fact that $\vec{\hat{k}}_{ab}$ is a Hodge dual of closed form $\vec{\hat{h}}_{ab}$ we automatically get
\begin{align}
    \vec{\varsigma}[\vec{\hat{k}}]_{b} &=0\,.
\end{align}

A conformal Killing tensor (CKT) is a square of CKY tensor. 
It can be checked that $\vec{K}_{ab}=\vec{k}_a^{\phantom{a}c}\vec{k}_{cb}$ is a proper Killing tensor (KT).
The remaining Killing vector is constructed from Killing tensor  by contraction with the Killing vector $\vec{\KVxi}_a$ and it reads
\begin{align}
    \vec{\tilde{\zeta}}^a &=\vec{K}^{ab}\vec{{\tilde{\xi}_{(t)}}}_b \\
    &=-\frac{1}{4}\kappa_1 \Bigl[
    (\kappa_1-\bar{\kappa}_{1'})^2\GHPrho'\,\vec{l}^a
    -(\kappa_1-\bar{\kappa}_{1'})^2\GHPrho\,\vec{n}^a \nonumber\\
    &\phantom{-\frac{1}{4}\kappa_1 \Bigl[}-(\kappa_1+\bar{\kappa}_{1'})^2\tau'\,\vec{m}^a
    +(\kappa_1+\bar{\kappa}_{1'})^2\tau\,\vec{\bar{m}}^a\Bigr]\nonumber\\
    \vec{\zeta}^a &= \frac{1}{2}\left( \vec{\tilde{\zeta}}^a+\vec{\bar{\tilde{\zeta}}}^a\right)=a^2\vec{\p_t}^a+a\vec{\p_\phi}^a\,.
\end{align}
Again, if $\vec{\hat{h}}_{ab}$ is principal KY already $\vec{\tilde{\zeta}}^a$ is real vector.

We will follow in a modified fashion \cite{edgar_petrov_2009} and derive post Bianchi identities by applying GHP commutators on $\kappa_1$, we get:
\begin{align}
    \thorn\GHPrho' &= \thorn'\GHPrho+\tau\bar{\tau}-\tau'\bar{\tau}'\,, & 
    \thorn'\tau&=\eth'\GHPrho\,.
    \\
    \eth\tau' &=\eth'\tau+\GHPrho\bar{\GHPrho}'-\bar{\GHPrho}\GHPrho'\,, &
    \eth \GHPrho'&=\thorn\tau'
\end{align}

In the next step we follow \cite{price_developmenst_2015}. We have two independent KVs (possible KTs constructed from these are divergence free) one irreducible KT $\vec{K}_{ab}$ (thus divergence-free) and manifestly trace-free CKY
\begin{align}
    \vec{P}_{ab}&=\kappa_{AB}\bar{\kappa}_{A'B'}=2\kappa_1\bar{\kappa}_{1'}
    \left( \vec{l}_{(a}\vec{n}_{b)}+\vec{m}_{(a}\vec{\bar{m}}_{b)}\right).
\end{align}
The KT can be split up to trace-less part and trace as
\begin{align}
    \vec{K}_{ab} &= \vec{P}_{ab}+\frac{1}{4}K\vec{g}_{ab}\,.
\end{align}
The projections of divergence of this equation gives 4 equations for GHP derivatives of $K$
\begin{align}
    \thorn K &= +4\kappa_1\bar{\kappa}_{1'}(\GHPrho+\bar{\GHPrho})\,, &
    \thorn' K &=+ 4\kappa_1\bar{\kappa}_{1'}(\GHPrho'+\bar{\GHPrho}')\,,\\
    \eth K &= -4\kappa_1\bar{\kappa}_{1'}(\tau+\bar{\tau}')\,,&
    \eth' K &= -4\kappa_1\bar{\kappa}_{1'}(\bar{\tau}+\tau')
\end{align}
The integrability condition then result in next set of identities as follows
\begin{align}
    \tau\bar{\tau} &= \tau'\bar{\tau}'\,,  &
    \thorn\tau' &= 2(\GHPrho\bar{\tau}+\GHPrho\tau'+\bar{\GHPrho}\tau')\,,\\
    \GHPrho\bar{\GHPrho}' &= \bar{\GHPrho}\GHPrho'\,, &
    \thorn'\tau &= 2(\GHPrho'\bar{\tau}'+\GHPrho'\tau+\bar{\GHPrho}'\tau)\,.
\end{align}
The last set of identities we need is recovered by checking the Killing vector equation for $\vec{\tilde{\zeta}}$ or applying commutators on $\GHPrho$ and reads
\begin{align}
    \GHPrho\bar{\tau}+\bar{\GHPrho}\tau' &=0\,, &\bar{\GHPrho}\tau+\GHPrho\bar{\tau}' &=0\,,\\
    \GHPrho'\bar{\tau}'+\bar{\GHPrho}'\tau &=0\,, &\bar{\GHPrho}'\tau'+\GHPrho'\bar{\tau} &=0\,.
\end{align}

\bibliography{kofron}

\end{document}